\documentclass[10pt,
aps, 
prd,
superscriptaddress,
showpacs,
showkeys,
notitlepage,
twocolumn, 
numbers,
longbibliography,
nofootinbib]
{revtex4-1}

\usepackage[T1]{fontenc} 
\usepackage[utf8]{inputenc} 
\usepackage[english]{babel} 
\usepackage{graphicx, xcolor}
\usepackage{amsfonts, amssymb, amsmath, mathrsfs, calc, array, mathdesign, mathtools}
\usepackage[separate-uncertainty=true]{siunitx}
\usepackage{hyperref}
\makeatletter
\newcommand{\raisedhypertarget}[1]{%
  \Hy@raisedlink{%
    \hyper@anchorstart{#1}\hyper@anchorend
  }%
}
\makeatother
\usepackage{abraces}

\usepackage{dsfont}
\usepackage{slashed}

\newcommand{\fsl}[1]{{\ooalign{\(#1\)\cr\hidewidth\(/\)\hidewidth\cr}}}

\usepackage{tikz}

\DeclareRobustCommand{\doublezigzag}{%
  \tikz[baseline=-0.5pt,x=0.18em,y=0.18em]{
    \draw[line width=0.08em,line cap=round]
      (0,4)--(3,4)--(0,2)--(3,2)--(0,0)--(3,0);
  }%
}
\usepackage{tabstackengine}
\stackMath
\newcolumntype{C}[1]{>{\centering\arraybackslash$}m{#1}<{$}}
   
\def\bra#1{\mathinner{\langle{#1}|}}
\def\ket#1{\mathinner{|{#1}\rangle}}

\def\ontop#1#2{\setbox0\hbox{#2}\copy0\llap{\raise\ht0\hbox{#1}}}
\definecolor{darkblue}{rgb}{0,0,0.93} 
\definecolor{darkred}{rgb}{0.8,0,0} 
\newcommand{\e}{\epsilon}
\newcommand{\defeq}{\vcentcolon=}
\newcommand{\defeqback}{=\vcentcolon}

\usepackage{esint}
\newcommand{\PArrighi}[1]{{\color{magenta} #1}}
\newcommand{\DB}[1]{{\color{blue} #1 }}

\hypersetup{
colorlinks=true, 
linkcolor=darkblue, 
citecolor=darkred, 
filecolor=darkblue, 
urlcolor=darkblue
}

\newif\ifusebibfile

\begin{document}

\title{Fermion-doubling problem in Chiral discretizations of Quantum field theory: \\ Definitive proof, Fixing, and Computation of two-point correlation functions}

\author{Dogukan Bakircioglu}
\email{dogukan.bakircioglu@universite-paris-saclay.fr}
\affiliation{Universit{\'e} Paris-Saclay, INRIA, CNRS, ENS Paris-Saclay, LMF, 91190 Gif-sur-Yvette, France}

\author{Pablo Arnault}
\email{pablo.arnault@ens-paris-saclay.fr}
\affiliation{Universit{\'e} Paris-Saclay, INRIA, CNRS, ENS Paris-Saclay, LMF, 91190 Gif-sur-Yvette, France}

\begin{abstract}

\noindent
We give the definitive proof that the Dirac Quantum Cellular Automaton (QCA) used for both quantum simulation and algorithmic foundations of Quantum Field Theory (QFT), and especially of Quantum Electrodynamics (QED), as put forward in Refs.\ \cite{ABF20,EDMMplus2023}, does exhibit Fermion \mbox{Doubling (FD)}, albeit thrice as less severe as discrete-time standard Lattice Gauge \mbox{Theories (LGTs)---as shown} in Ref.\ \cite{DAA2025}---, which are naive regarding the spacetime discretization of differential operators acting on fermionic fields. The proof is done for the $(1+1)$D Dirac-QCA model. We show that \mbox{the (one-time-step) two-point} correlation function, also called Green's \mbox{Function (GF), of} the Dirac QCA, is of astonishing simplicity, which is in contrast with the Green's function of the Dirac equation. We also compare, both qualitatively and quantitatively, this Dirac QCA to the continuous-time-LGT spatial discretization of Dirac fermions regarding how well these two lattice models approximate their naive continuum limit---which is nothing but the Dirac equation---even when very far away from this continuum limit, a situation which must be considered because of experimental limitations in quantum simulation---: the Dirac QCA is better for ultrarelativistic regimes, whereas continuous-time LGT is better for non-relativistic regimes. In a second part of this work, we compute the two-point correlation function of the FD-fixed model put forward in \mbox{Ref.\ \cite{DAA2025}}, called Flavored Dirac QCA (FDQCA)---which staggers an extra, artificial flavor \emph{only}, on a diamond spacetime lattice, and does not stagger chirality as staggered fermions in usual LGT. The structure of this FDQCA two-point correlation function is of extreme simplicity, and can be expressed in a very simple manner in terms of the four chiral components of the FD-suffering, original-model Green's function.

\end{abstract}



\maketitle

\section{Introduction}

\noindent
Quantum cellular automata (QCAs) are quantum-computing models on spacetime lattices---rather than on a multiwire computational basis having no physical significance---, which are natively purely local---i.e., no need for one- an two-qudit gate decompositions of the circuit---, and which also allow for the use of spatial translations, which may depend on each qudit-basis element. The discussion of how to decompose the spatial translations into one- and two-qudit gates is one of the major topics of the field of QCAs, but will not be the topic of the present article, which will simply make use of the aforementioned spatial translations as they are. 

In the present article, we use QCAs will be used for modelling the motions of several quantum particles, possibly interacting between each other---rather than, for example, modelling the flow of information into some Grover-like algorithm on a $2$D spatial grid. As any multiparticle quantum-mechanical model, QCAs for modelling particles admit a representation in terms of quantum fields, be they merely formal and just an alternate language as in non-relativistic QM, or a fundamental framework having no ``natural'' or at least simple equivalent formulation just with wavefunctions, as in quantum field theory (QFT).

In this work, we will use QCAs to model discrete-spacetime QFTs. Actually, QCAs are a novel type of lattice field theories (LFTs), which in addition to relativistic locality, which LFTs also have, enjoy, in addition, native, gauranteed unitarity of the evolution, and the possibility to respect the chirally dependent transport of certain continuum QFT fundamental equations such as no less than the Dirac equation, whereas in standard LFTs the latter is discretized naively, i.e., via finite differences which are independent of any internal indices/parameters of the quantum field\footnote{Recall the degrees of freedom of wavefunction quantum-mechanical model formulations, be they external or internal, become mere \emph{parameters} of the model when formulated in terms of quantum fields.}, which are always in finite and small number, such as chirality, spin, or gauge degrees of freedom.

In this article, the QFT model we will deal with is very simple: $(1+1)$D quantum electrodynamics (QED). In \mbox{Ref.\ \cite{DAA2025}}, we have put forward a minimal way of fixing the (specific type of) fermion-doubling (FD) problem exhibited by the QED-QCA seminal models put forward in Refs.\ \cite{ABF20, EDMMplus2023}. The FD problem is a model that affects quantum field theories containing interacting \emph{fermionic} fields. In \mbox{Ref.\ \cite{DAA2025}}, we show the problem is less severe in the QED-QCA lattice gauge theories of the previously mentioned two references than in standard lattice gauge theories (LGTs). This is directly due to the use of chirality-dependent spatial shifts in the two previous QCA models, rather than naive (symmetric) finite differences in standard LGTs. Very recently, \mbox{Ref.\ \cite{bakircioglu2026flavouredlatticeschwingermodel}} employed the \mbox{FD-fixing} method developed in \mbox{Ref.\ \cite{DAA2025}} in the lattice Hamiltonian formulation of the Schwinger \mbox{model---formulation} which is in discrete space but continuous time---, demonstrating a realization of the chiral anomaly on the lattice.

Although the FD problem only affects \emph{interacting} field theories (containing fermions)---which anyways are usually the only interesting ones---, the problem can usually be fixed, not only on the mere \emph{free} theory, but on the mere (mathematically) classical version of the field theory, and this is what is done in \mbox{Ref.\ \cite{DAA2025}}. Of course, the mathematically classical version of the \emph{fermion} quantum field of the model, has nothing classical physically speaking: it is, rather, a (single-particle) quantum-mechanical wavefunction. More precisely, it is a Dirac field. Now, restricting ourselves to the classical version of the theory to study and fix FD also means, in a way, restricting ourselves to the single-particle sector of the theory. We wish to mention that, while formally the previous-sentence remark will not pose any particular question in the quantum-field-formalism formulation of the QCA, in the qubit or gate formalism, this does mean restricting oneself to only one linear sector/subspace of the dynamics. 

Before continuing on fermion doubling, let us give some information about the free lattice fermion models we are dealing with; this information is recalled in detail in \mbox{Sec.\ \ref{sec:Dirac QCA}}. First of all, in this article we restrict ourselves to the $(1+1)$D QED QCA of \mbox{Ref.\ \cite{ABF20}---i.e., we} will not treat the $(3+1)$D model of \mbox{Ref.\ \cite{EDMMplus2023} (and} not its $(2+1)$D model either). Now, the free fermion part of that model is a particular Dirac QCA, where we define a Dirac QCA simply as a QCA whose \emph{naive} continuum limit delivers the Dirac equation. Finally, the mathematically classical version of this (and of any) Dirac QCA, which is also equivalent to its single-particle sector, is known to be a so-called Dirac discrete-time quantum walk (DQW), which we will often simply write Dirac QW. It is on the Dirac-QW model that all the FD analysis and fixing will be done.

So let us come back to this FD problem. The simplest way to identify some FD problem in a spacetime lattice model with fermionic quantum fields, is by looking at the poles of the spacetime-Fourier-space fermion Green's function (GF) of the model---a quantity computable on the mere free model---, which often simply correspond to the zeroes of the determinant of the spacetime-Fourier-space fermion equation-of-motion (EOM) operator of the single-particle free model---here, the Dirac QW. This spacetime-Fourier-space GF (FGF) of the model is by construction the matrix inverse (times $i$ with our choice of convention) of the EOM operator just \mbox{mentioned---hence} the fact that we have to look at the aforementioned zeroes. At this point, let us mention as a vocabulary note that the 

Now, while the identification of these zeroes has already been done in \mbox{Ref.\ \cite{DAA2025}, in} that article we have not \emph{explicitly} shown that these zeroes indeed contribute to spurious doublers, by actually performing any of either the momentum or of the energy integral of the GF. In this article we do so, in Sec.\ \ref{sec:correlationfunctionQCA}, on the single-time-step GF, via a first contour integration to perform the energy integral---as it is most standard and natural, since the result yields the sole momentum Fourier integral of an object which is necessarily the momentum Fourier transform of the GF as a function of the transition \mbox{distance $x-x' \in \mathbb Z \e$, and} this whole momentum integral can be identified to what one obtains when one inserts a momentum closure relation in the standard quantum-mechanical propagator. It is hence shown that all the poles expected to be problematic indeed contribute to spurious doublers. 
Moreover, we also comment on the shape of the fully spacetime-Fourier-integrated FGF: essentially, it is a closed form, which in addition is extremely simple since, when it comes to the spatial degree of freedom, it involves only Kronecker deltas between positions, and even for the chirality degree of freedom, we simply need to know the coefficients $\cos (\e m)$ and $\sin (\e m)$. This simplicity is in contrast with the form of the Dirac-equation's Green function, which, first of all, is not even a closed form, and must be expressed in terms of Hankel or of Bessel functions.

Now, one first thing we would like to stress at this point is that, now that we know that the Dirac QW suffers from a FD problem, one must always systematically have in mind that there are always two ``meanings'' of the wording ``continuum limit'': there is (i) the well-known \emph{naive} continuum limit, which is ``blind'' to the FD problem, and there is (ii) the \emph{true}, i.e., \emph{correct} continuum limit, which takes into account the FD problem, and which, hence does \emph{NOT} coincide with the naive continuum limit, which is the standard Dirac equation here. This, along the way, should remind, or make the reader realize how big this FD problem is.

In a ``second part'' of Sec.\ \ref{sec:correlationfunctionQCA}, we compare two different lattice models between them, as compared to their joint naive continuum limit, namely, the Dirac equation. These two lattice models, are, first, of course, the Dirac QW, and second, continuous-time naive lattice fermions---from continuous-time LGT---, which attracts much interest in the community of analog and even digital quantum simulation. Of course, as we have mentioned, both these models have as naive continuum limit the Dirac equation. That being said, in quantum simulation experiments, the continuum-limit regime is difficult to reach because of experimental limitations, and hence we propose a study of how well the Dirac QW and continuous-time naive lattice fermions \emph{approximate} the Dirac equation, even when they cannot be considered close to this continuum limit. We find out, quantitatively, in which sense one can say that (i) the Dirac QW approximates the Dirac equation better than continuous-time naive lattice fermions in the ultrarelativistic regime which we define, and that (ii) it is the converse that happens in the non-relativistic regime which we define.

In a second part of this article, constituted by \mbox{Sec.\ \ref{sec:correlationFQCA}}, we compute the GF of the FD-fixed QCA model put forward in \mbox{Ref.\ \cite{DAA2025}, which} we call $G^{\text{FQW}}(t,x;t',x')$. More precisely, we provide a fully explicit expression in terms of the $4$ chirality-space components of the FD-non-fixed model, that is, $G_{ij}^{\text{QW}}(t,x;t',x')$ for $i,j=1,2$, where $i=1$ corresponds to chirality ``$+$'' (wavefunction going to the right in time on the standardly, right-positively-oriented $1$D spatial space), and $i=2$ corresponds to chirality ``$-$''.

In Sec.\ \ref{sec:lastsection}, we give some conclusions and discuss our results.

\section{Dirac-QCA model in $1+1$ dimensions}
\label{sec:Dirac QCA}

\subsection{Direct-space scheme}

\noindent
The starting point of this work is the specific $(1+1)$D Dirac QCA defined in \mbox{Ref.\ \cite{ABF20}. A standard} way of writing the one-particle-sector dynamics of this Dirac QCA is the following,
\begin{equation}
\label{eq:themodel}
\psi(t+\epsilon,x) = \mathcal U^{\text{QW}}_{\epsilon} \psi(t,\cdot) \big|_{x} \, .
\end{equation}
This equation is an evolution equation by one time step $\epsilon \in \mathbb{R}_+$, for a single-particle wavefunction $\psi : (t,x) \mapsto \psi(t,x)$ from $(\mathbb Z \epsilon)^2 $ to $\mathbb C^2$, via a unitary one-time-step evolution operator $\mathcal U^{\text{QW}}_{\epsilon}$. The time step $\epsilon$ is also a space step, since we take natural units. The unitary one-time-step evolution operator $\mathcal U^{\text{QW}}_{\epsilon}$ is defined by
\begin{equation}
\mathcal U^{\text{QW}}_{\epsilon} \defeq \begin{bmatrix}
				\cos(m \epsilon) \mathcal{S}_{\epsilon} &  - i \sin(m \epsilon) \\
				- i \sin(m \epsilon) & \cos(m \epsilon) \mathcal{S}^{\dagger}_{\epsilon}
			\end{bmatrix} \, .     
\end{equation}
In the previous expression, we have introduced (i) the spatial shift operator $\mathcal S_\epsilon$, that can be defined via its action on an arbitrary wavefunction $\phi: (t,x) \mapsto \phi(t,x)$, namely,
\begin{equation}
\mathcal S_{\epsilon} \phi(t)\big|_x \defeq \phi(t,x-\epsilon) \, ,
\end{equation}
and (ii) the mass $m$ of the spin-$1/2$ ``classical'' Dirac field\footnote{We will always put quotation marks around the word ``classical'' in such a context, because this word is extremely confusing for scientific communities other than that of physical field theories. Indeed, this word ``classical'' between quotations marks must be understood, for $\mathbb C$-valued such fields, in a \emph{purely} mathematical sense (i.e., it means the field is not a field operator, i.e., not an operator-valued field, but merely a number-valued field), not in a physical one, since physically for example a ``classical'' Dirac field has nothing classical, physically its is a purely \emph{quantum} object.} that is obtained in the naive continuum limit $\epsilon \rightarrow 0$ of the model, see \mbox{App.\ \ref{app:contlim1}}. The operator $\mathcal S^\dagger_{\epsilon}$ is the adjoint, i.e., the Hermitian conjugate of $\mathcal S_{\epsilon}$, which can be shown to act as
\begin{equation}
\mathcal S^\dagger_{\epsilon} \phi(t,\cdot)\big|_x = \phi(t,x+\epsilon) \, .
\end{equation}
The superscript ``QW'' stands for ``quantum walk'', since it is well known that the one-particle sector of QCAs corresponds to discrete-time quantum walks (DQWs) \cite{Vogts09}. In the present context, we call our DQW a Dirac DQW.

We must now mention two particularities of the context in which we have carried out the present work. First, this work (i) is anyways partially related to multiparticle quantum mechanics and especially quantum field theory\footnote{We say ``partially'' because we remain in the one-particle sector of the QCA, since this is enough in the case of the \emph{two-point} correlation function \emph{for a free model}---and we know that fixing some FD problem, as part of the topic of this article, is always done on the free model, since this is enough.}, and, more than that, (ii) it is bound to be of much use in these contexts, so that it makes sense to use right away a formalism that is close to these contexts. Second, our discretization of the Dirac equation is in both space \emph{and} time, which implies that, in order to treat its FD problem\footnote{In the FD vocabulary, established seminally by the LGT scientific community, we sometimes say that, when time is discrete, we may have \emph{temporal} doublers in addition to the \emph{spatial} doublers already appearing for example in \mbox{continuous-time LGT---Ref.\ \cite{Alford1997}}, for example, uses this vocabulary. (``Pure'') spatial doublers also appear in naive space \emph{and} time discretizations of continuum fermionic theories involving a chirality degree of freedom. In QCA, however, there are no spatial doublers, nor, actually, temporal doublers either; there are only what we have called \emph{spatiotemporal} doublers \cite{DAA2025}. These spatiotemporal doublers, as the (``purely'') temporal ones, only potentially arise when time is discrete.}~\cite{DAA2025}, it is convenient to treat all relevant dynamical objects on the \emph{whole} (discrete) spacetime whenever possible---and not only on the purely ``spatial space'' as in continuous-time LGT. 

Hence, we will write the one-particle-sector dynamics of the $(1+1)$D Dirac QCA as follows,
\begin{equation}
\label{eq:the_eq}
\mathcal M^{\text{QW}}_{\epsilon} \psi |_{t,x} = 0 \, ,
\end{equation}
where we have introduced the equation-of-motion (EOM) operator 
\begin{equation}
\mathcal M^{\text{QW}}_{\epsilon} \defeq  \mathcal U^{\text{QW}}_{\epsilon} -  \mathcal T^{\dagger}_{\epsilon} \, ,
\label{eq:eomop}
\end{equation}
where $\mathcal T_{\epsilon}$ is the temporal shift operator, that can be defined via its action on an arbitrary wavefunction $\phi$, namely,
\begin{equation}
\mathcal T_{\epsilon} \phi(\cdot,x)\big|_t \defeq \phi(t-\epsilon,x) \, .  
\end{equation}
The operator $\mathcal T^\dagger_{\epsilon}$ is the adjoint of $\mathcal T_{\epsilon}$, which can be shown to act as
\begin{equation}
\mathcal T^\dagger_{\epsilon} \phi(\cdot,x)\big|_t = \phi(t+\epsilon,x) \, .
\end{equation}
Basically, we have treated time on the same footing as space, and the time instant $t$ and the position $x$ are \emph{parameters} of the ``classical'' \mbox{field---rather} than the position being a \emph{dynamical variable} bound to be quantized as in standard quantum mechanics.

For the sake of completeness, let us write the EOM operator of Eq.\ \eqref{eq:eomop} explicitly:
\begin{equation}
\mathcal M^{\text{QW}}_{\epsilon} = 
    \begin{bmatrix}
        c_{\epsilon} \mathcal S_{\epsilon} - \mathcal T_{\epsilon}^\dagger & -i s_{\epsilon}\\
       - i s_{\epsilon} &   c_{\epsilon} \mathcal S_{\epsilon}^{\dagger} -\mathcal T_{\epsilon}^\dagger
    \end{bmatrix}
 \, ,
\label{eq:eomop2}
\end{equation}
where $c_\epsilon \defeq \cos (m \epsilon)$ and $s_\epsilon \defeq \sin (m \epsilon)$. \\

\subsection{Fourier-space scheme}

\noindent
We are going to look for solutions $\psi(t,x)$ of Eq.\ \eqref{eq:the_eq} under the form of the following decomposition ansatz,
\begin{equation}
\label{eq:FT}
\psi(t,x) = \int_{- \pi/\epsilon}^{\pi/\epsilon}  \frac{dE}{2 \pi } \int_{- \pi/\epsilon}^{\pi/\epsilon}  \frac{dp}{2 \pi } \hspace{1mm}  e^{-iEt + ipx}  \tilde \psi(E,p) \, ,
\end{equation}
where the $\tilde \psi (E,p)$'s are the coefficients of this decomposition, coefficients to be determined, and which can be seen as the following formal spacetime Fourier transform:
\begin{equation}
\tilde \psi(E,p) = \epsilon^2 \sum_{t \in \mathbb{Z}\epsilon} \sum_{x \in \mathbb{Z}\epsilon} e^{iEt - ipx}  \psi(t,x) \, .
\end{equation}
Inserting Eq.\ \eqref{eq:FT} into Eq.\ \eqref{eq:the_eq}, we finally obtain
{\small
\begin{equation}
 \int_{- \pi/\epsilon}^{\pi/\epsilon}  \frac{dE}{2 \pi } \int_{- \pi/\epsilon}^{\pi/\epsilon}  \frac{dp}{2 \pi } \hspace{1mm}  e^{-iEt + ipx} M^{\mathrm{QW};\mathcal B}_\epsilon (E,p) \tilde \psi(E,p)  = 0 \, ,
\end{equation}}%
where
\begin{equation}
M^{\mathrm{QW};\mathcal B}_\epsilon(E,p) \defeq 
\begin{bmatrix}
c_\epsilon e^{-i\epsilon p} - e^{-i \epsilon E} & - i s_\epsilon \\
- i s_\epsilon & c_\epsilon e^{i\epsilon p} - e^{-i \epsilon E}
\end{bmatrix} \, ,
\end{equation} 
which we call Fourier-space EOM operator, and where $\mathcal B \defeq \{(p,E)\in [-\pi/\epsilon,\pi/\epsilon]^2\}$ is the spacetime Brillouin zone (BZ). Of course, since $E$ is the Fourier variable associated to $t$, it physically represents the energy of the field, and since $p$ is the Fourier variable associated to $x$, it represents the linear momentum in the $x$ direction.

\section{Correlation function of the Dirac QCA: its simplicity, but also its spurious fermion doubling (FD)}
\label{sec:correlationfunctionQCA}

\subsection{Green's function: usefulness and definition}

\noindent
{\bfseries Usefulness.}
The basic usefulness of Green's function $G(t,x; t',x')$ in mathematically classical field \mbox{theory---where} the most desirable aim is usually to find a solution of the EOM, and then to compute observable quantities out of it\footnote{When finding a solution to the EOM is too difficult, we may try to compute the desired observable directly, without having solved the EOM.}---is that it enables to evolve an arbitrary initial state (starting at, say, $t'$) up to any final time $t$, via the following equation---which is very standard, but adapted to the present discrete-spacetime context---,
\begin{equation}
\label{eq:propagationequation}
 \psi(t,x) =   \sum_{x' \in \mathbb Z \epsilon} \epsilon \, G(t,x; t',x') \psi(t',x') \, .
\end{equation}

\noindent
{\bfseries Definition.} It turns out that, in our framework, Green's function has to be defined as the ``inverse'', not of the EOM operator that we have already introduced---both for pedagogy and for ``historical'' reasons---in \mbox{Eq.\ \eqref{eq:eomop}}, but of the following modified version of it,
\begin{subequations}
\begin{align}
\mathcal{L}^{\text{QW}}_{ \epsilon} &\defeq \frac{1}{i \epsilon} \mathcal T_\epsilon \mathcal{M}^{\text{QW}}_{\epsilon} \\
&=  \frac{1}{i \epsilon} (\mathcal T_\epsilon \mathcal U_\epsilon^{\text{QW}} - \mathds 1^{\text{QW}}) \label{eq:modifiedEOMoperator} \, , 
\end{align}
\end{subequations}
where $\mathds 1^{\text{QW}} $ is the identity operator acting on the Hilbert space to which $\psi$ belongs.

We now define Green's function $G(t,x; t',x')$ as the solution of the following equation\footnote{If $\mathcal L_{\epsilon}^{\text{QW}}$ was $\mathcal L^{\text{Dirac}}$, the linear operator of the standard Dirac equation in Lorentz-covariant form, the following convention used for the definition of the Green's function is \emph{not} that used in Schwartz's QFT book \cite{book_Schwartz}: in the latter there is no $i$---see \mbox{Eq.\ (18.4)} of that book for example. But the convention we use is often used in non-relativistic quantum mechanics---see for example \mbox{Ref.\ \cite{Littlejohn}---, with} which we want to connect because QCAs have a clear time arrow, and are much closer, in their formalism---this is of course related to them being a model of quantum computation---, to standard, i.e., Hamiltonian quantum mechanics. In relation to what we have just said, in \mbox{App.\ \ref{app:contlim1}} we actually show that the naive continuum limit of $\mathcal L_{\epsilon}^{\text{QW}}$ is simply $\mathcal L_{\epsilon}^{\text{Ham.\ Dirac}}$, the Hamiltonian, or Schrödinger-type-equation version of $\mathcal L^{\text{Dirac}}$.},
\begin{equation}
\label{eq:thedefeq}
\mathcal{L}^{\text{QW}}_{\epsilon} G(\cdot,\cdot;t',x') \big|_{t,x} = i \left(\frac{1}{\epsilon}\delta_{t,t'}\right) \left(\frac{1}{\epsilon}\delta_{x,x'}\right)  \, ,
\end{equation}

Notice that in \mbox{Ref.\ \cite{DAA2025}, we} defined Green's function as the ``inverse'' of $\mathcal{M}^{\text{QW}}_{\epsilon}$ rather than of $\mathcal{L}^{\text{QW}}_{\epsilon}$: this is not the best definition, the best one is the one of the present article, that is, the ``inverse'' of $\mathcal{L}^{\text{QW}}_{\epsilon}$. Nevertheless, we will use, in this article, essentially the same notations as in \mbox{Ref.\ \cite{DAA2025}, which} are the simplest, so we ask the reader to have in mind this difference between the Green's function defined in the present article, which is the best one, and that of \mbox{Ref.\ \cite{DAA2025}}, in particular since we have used essentially the same notation as the latter for the former, that is, the sole \mbox{letter $G$}.

\subsection{Computation of Green's function (GF)}

\noindent
{\bf Fourier-space GF for the $(1+1)$D Dirac QCA.}
We are going to look for a solution of the \mbox{Green's-function} definition equation under the form of the following ansatz,
\par\nobreak
{\small
    \begin{equation}\label{eq:propagator}
	G(t,x ; t',x') =  \int_{- \pi/\epsilon}^{\pi/\epsilon}  \frac{dE}{2 \pi } \int_{- \pi/\epsilon}^{\pi/\epsilon}  \frac{dp}{2 \pi } \hspace{1mm}  
			e^{-iE(t-t')+ ip(x-x')}  G_{\mathcal{B}}(E,p) \, ,
    \end{equation}}
where $G_{\mathcal{B}}(E,p)$ can be seen, formally, as the spacetime Fourier transform of $G(t,x ; t',x')$. This is standard in Green's function theory, as the ``philosophy'' of this theory is that it is simpler to invert the Green's-function definition equation in Fourier space, where linear operators are replaced by mere numbers.

Using the method recalled in Ref.\ \cite{DAA2025}, which starts from the fact that one actually simply has
\begin{equation}
\label{eq:inverting}
G_{\mathcal B}(E,p) = i (L^{\text{QW};\mathcal B}_{\epsilon}(E,p))^{-1} \, ,
\end{equation}
where the definition of $L^{\text{QW};\mathcal B}_{\epsilon}(E,p)$ is analogous to that of $M^{\text{QW};\mathcal B}_{\epsilon}(E,p)$, we find, after a few lines of computation,
\par\nobreak
{\small
\begin{equation}
\label{eq:final_expression}
G_{\mathcal B}(E,p) = \frac{1}{ \epsilon \, \mathscr D(E,p))}
\begin{bmatrix}
c_\epsilon e^{i \epsilon E} e^{i\epsilon p} - 1 &  i s_\epsilon e^{i \epsilon E} \\
 i s_\epsilon e^{i \epsilon E} & c_\epsilon e^{i \epsilon E} e^{-i\epsilon p} - 1
\end{bmatrix}  \, ,
\end{equation}}
where
\begin{subequations}
\begin{align}
\mathscr D(E,p) &\defeq \det \{ L^{\text{QW};\mathcal B}_{\epsilon}(E,p) \} \label{eq:LB}\\
&= \det \{ \frac{1}{i \epsilon} e^{i\epsilon E} M^{\text{QW};\mathcal B}_{\epsilon}(E,p) \} \\
&= - \frac{1}{\epsilon^2}  e^{2i\epsilon E} \det \{ M^{\text{QW};\mathcal B}_{\epsilon}(E,p) \} \\
&= - \frac{1}{\epsilon^2} \underbrace{\big( e^{2i \epsilon E } - 2 c_\epsilon \cos(p \epsilon) e^{i \epsilon E } +1 \big)}_{\defeqback \, \mathscr D^{\text{bis}}(E,p)} \, ,
\end{align}
\end{subequations}
which, inserted in Eq.\ \eqref{eq:final_expression}, yields
\par\nobreak
{\small
\begin{align}
\label{eq:final_expression_bis0}
G_{\mathcal B}(E,p) &= - \frac{\epsilon}{\mathscr D^{\text{bis}}(E,p)}
\begin{bmatrix}
c_\epsilon e^{i \epsilon E} e^{i\epsilon p} - 1 &  i s_\epsilon e^{i \epsilon E} \\
 i s_\epsilon e^{i \epsilon E} & c_\epsilon e^{i \epsilon E} e^{-i\epsilon p} - 1
\end{bmatrix}  \, .
\end{align}}

\noindent
{\bf Direct-space one-time-step GF.} In order to obtain the direct-space GF, we go back to Eq.\ \eqref{eq:propagator}, and insert in it the expression obtained for the Fourier-space GF in Eq.\ \eqref{eq:final_expression_bis0}. In \mbox{App.\ \ref{app:Dirac-QCA_GF}}, we show that this eventually leads to the following expression if we choose $t-t'=\epsilon$:
\par\nobreak
{\small
\begin{equation}
\label{eq:FFinal}
G(t'+\epsilon,x;t',x') = \int_{-\pi/\epsilon}^{\pi/\epsilon}  \frac{dp}{2\pi} e^{i p (x-x')}
\begin{bmatrix}
  c_{\epsilon} e^{-i\epsilon p} & - i s_{\epsilon} \\
 - i s_{\epsilon} &   c_{\epsilon} e^{i\epsilon p} 
\end{bmatrix} \, .
\end{equation}}
In App.\ \ref{app:propagator}, we show that the RHS of the previous equation corresponds to the propagator of our Dirac DQW, defined by
\begin{equation}
 K(t'+\epsilon,x;t',x') \defeq \langle x | \hat{U}_\epsilon^{\text{QW}} | x' \rangle \, ,
\end{equation}    
where $\hat{U}_\epsilon^{\text{QW}}$ is the abstract-QM version of $\mathcal{U}_\epsilon^{\text{QW}}$; this ``correspondance'' is nothing but an equality:
\begin{equation}
 G(t'+\epsilon,x;t',x') =  K(t'+\epsilon,x;t',x') \, .  
\end{equation}

Luckily, the expression of Eq.\ \eqref{eq:FFinal} is integrable into a closed form, and we show in App.\ \ref{app:Finalexpr} that this yields
\begin{equation}
\label{eq:thefinaleqisthisone}
G(t'+\epsilon,x;t',x') = \frac{1}{\epsilon}\left[\begin{array}{cc}
 c_\epsilon \delta_{x',x-\epsilon} & - i s_\epsilon \delta_{x',x}  \\
 - i s_\epsilon \delta_{x',x}  &  c_\epsilon \delta_{x',x+\epsilon}
\end{array}  \right]  \, .
\end{equation} 
In App.\ \ref{app:TaylorExpansion}, we perform the Taylor expansion of the previous expression in $\epsilon$, and we show that inserting this Taylor expansion in the ``propagation equation'', \mbox{Eq.\ \eqref{eq:propagationequation}}, indeed delivers the Dirac equation. \\

\noindent
{\bf Comments.} We would like to draw the reader's attention to the following facts. As we have just evoked, the previous expression is a closed form\footnote{Standardly, a closed form is an analytical expression that is written in terms of the basic arithmetic operations ($+$, $-$, $\times$, and $/$) and of function composition involving only functions considered basic enough. Typically, the functions allowed are exponentials, logarithms, and trigonometric functions.}. This is in contrast, of course, with the continuum-model situation, since the Green's function of the Dirac equation is \emph{not} a closed form: indeed, that continuum-model Green's function can only be expressed in terms of Hankel or of Bessel functions \cite{Bernd_Thaller}.  

The fact that our Dirac-QCA Green's function is a closed form is of course, naively at least, \emph{a priori} an advantage with respect to the continuum-model situation, since this should allow for simpler computations\footnote{For example, one could already ask oneself whether the existence of such a closed form is an advantage in order to compute, or at least to express the generic $n$-step Green's function of our model, i.e., for an arbitrary time step $n \in \mathbb Z$.}.
From this \emph{a priori} computational advantage to saying that our Dirac-QCA is a better foundation of Dirac-fermion propagation in QFT, there is a gap, which we do not cross, but still we want to suggest this possibility. The main question one should treat here to position ourselves more conclusively with respect to the previous suggestion, is whether this closed form of the Green's function leads to simpler renormalization procedures in the interacting QFTs that one could attempt to ground on our free Dirac QCA---such as the models put forward in Refs.\ \cite{ABF20,EDMMplus2023}. What we have for sure---as in, however, any LGT---, is a natural ultraviolet regulator of the theory, namely, the spacetime-lattice spacing $\epsilon$. \emph{A priori}, however, it is actually difficult to say more than this, and anyways this ``this'' is actually not related to having a closed form for the Dirac-QCA Green's function. A question one could solve be solely analyzing free theories, is whether this closed form of our Dirac QCA is a feature specific to that lattice model, or if standard LGTs also have this feature: later in this article, we show that for example the continuous-time naive symmetric-finite-difference spatial discretization of the Dirac equation does not have this feature, so that our Dirac QCA seems particularly exceptional in this respect. When it comes to \emph{discrete}-time naive symmetric-finite-difference spatial discretization of the Dirac equation, one can actually show (unpublished yet) that these are not even unitary, so that let us say that with respect to that fact the closed-form question seems secondary at this stage.

\subsection{Witnessing the fermion-doubling problem: the two different meanings of the ``dispersion relation'' concept}
\label{subsec:witnessingFD}

\noindent
The magnitude of some potential fermion-doubling problem of a model ``M'' having Green's function $G^{\text{M}}(t,x;t',x')$ is witnessed, typically, by ``looking at'' the integrand of the space and time inverse-Fourier-transform expression for $G^{\text{M}}(t,x;t',x')$, which has the form
\par\nobreak
{\footnotesize
\begin{align}
 &G^{\text{M}}(t,x;t',x') \nonumber \\ 
&=  \int_{- E^{\text{max}}_\text{M}}^{E^{\text{max}}_\text{M}}  \frac{dE}{2 \pi } \int_{- p^{\text{max}}_\text{M}}^{p^{\text{max}}_\text{M}}  \frac{dp}{2 \pi } \hspace{1mm} e^{-iE(t-t')+ ip(x-x')}  G^{\text{M}}_{\mathcal{B}_\text{M}}(E,p)  \, , \label{eq:genericG} 
\end{align}}
and more precisely, it is $G^{\text{M}}_{\mathcal B_{\text{M}}}(E,p)$ that must be analyzed, and whose expression is
\begin{equation}
\label{eq:theinverseofG}
 G^{\text{M}}_{\mathcal B_{\text{M}}}(E,p) = i (L_\epsilon^{\text{M};\mathcal B_{\text{M}}}(E,p))^{-1} \, , 
\end{equation}
where $L_\epsilon^{\text{M};\mathcal B_{\text{M}}}(E,p)$ is the EOM operator of Model ``M''. Note that we have introduced the energy-momentum integration domain of Model ``M'': $\mathcal B_{\text{M}}$. Even more precisely, at this point one has to look at the dispersion relation. But we must distinguish here between two meanings of this terminology ``dispersion relation''.   \\

\noindent
{\bfseries First meaning.}  The standard basic relationship called ``dispersion relation'', is the vanishment of the determinant of the EOM Fourier-space operator, that is, here,
\begin{equation}
\label{eq:disprel}
 \det\{ L_\epsilon^{\text{QW};\mathcal B}(E,p) \} = 0 \, .   
\end{equation}
Now, this relation is important because we can see in Eq.\ \eqref{eq:theinverseofG} that $G^{\text{M}}_{\mathcal B_{\text{M}}}(E,p)$ is essentially the inverse \mbox{of $L_\epsilon^{\text{M};\mathcal B_{\text{M}}}(E,p)$,} and it is known, in relation to so-called Cramer's rule for solving linear systems of equations,  that the inverse of a matrix is given by its adjoint matrix \emph{divided by its determinant}. Hence, we are now in a position to understand that the zeros of $\det\{ L_\epsilon^{\text{QW};\mathcal B}(E,p) \}$, i.e., the solutions of the dispersion relation---understood as the previous equation---, or, in other words, the poles of $G^{\text{M}}_{\mathcal B_{\text{M}}}(E,p)$, will make big contributions to the integral of Eq.\ \eqref{eq:genericG}. This has already been said in Ref.\ \cite{DAA2025} \mbox{for $\text{M} = \text{QW}$,} but one of the points of this article is to shed as much light as possible on this. 

In the present article, indeed, it has been shown, in the case of $\text{M}=\text{QW}$, by doing a contour integration, see App.\ \ref{app:Dirac-QCA_GF}, that any couple $(E_i,p_i)$ that is a zero of the determinant $\mathscr{D}^{\text{alt.}}(E,p)$, will provide (this is implicit in our computation) a couple $(z_+,p_i)$ or $(z_-,p_i)$ that will be a zero of the ``complex-variable version'' of $\mathscr{D}^{\text{alt.}}(E,p)$, namely, $\Delta(z;p)$, and then it has been shown \emph{explicitly}, via the residue theorem, that this pole will \emph{for sure} contribute non-vanishingly to the integral over $E$. The same type of explicit proof can be done in general for any \mbox{model ``M'',} and may lead to poles of $G^{\text{M}}_{\mathcal B_{\text{M}}}(E,p)$ that will have significant contributions to the integral over $E$. In particular, this can be done for example for \emph{continous}-time so-called ``naive fermions'' of LGT, which we will denote by $\text{M}=\text{LGT}$. 

But what must be noticed here in particular is that, very generally speaking, other couples \mbox{than $(E_0=0,p_0=0)$ may} be zeros of the dispersion relation, whereas in the naive-continuum-limit model, only that couple is a zero, by construction. All other zeros are called doublers: they are spurious, because they always contribute to $G^{\text{M}}_{\mathcal B_{\text{M}}}(E,p)$, and if, after such a computation, usually via the residue theorem, one takes the continuum limit of the resulting expression for $G^{\text{M}}_{\mathcal B_{\text{M}}}(E,p)$---this is what we call a ``non-naive'' continuum limit---, this does \emph{not} remove the spurious contribution, yielding a mismatch with what one obtains if one directly computes the Fourier-space GF of the \emph{naive} continuum limit of Model \text{M}. \\

\noindent
{\bfseries Second meaning.} Let us now review the second meaning of the terminology ``dispersion relation''. To state it simply, this second meaning refers to the solutions  of the equation of the first meaning, i.e., \mbox{Eq.\ \eqref{eq:disprel}}, but not in terms of couples $(E_i,p_i)$ satisfying that equation, rather, as energy solutions being \emph{functions of the momentum}, i.e., of the form $f^{\text{M}}_b(p)$, $b\in \mathbb N$ being an energy-``branch'' index.
These energy branches are actually nothing but the eigen-energies or eigen-effective-energies of the expression being a function of $p$ that is obtained when performing the $E$-integral \mbox{$\int_{-\pi/\epsilon}^{\pi/\epsilon} (dp/(2\pi)) e^{-iE\epsilon}G_{\mathcal B}^{\text{M}}(E,p)$---so, for $t-t'=\epsilon$---,} expression which simply corresponds to the spatial Fourier transform of the usual quantum-mechanical propagator of Model $\text{M}$ considered as a function of $x-x'$. 

In the case of $\text{M}= \text{QW}$ \emph{versus}, or simply ``compared to'' $\text{M}= \text{LGT}$, the resulting energy branches are already well known, and recalled in Fig.\ 2 of Ref.\ \cite{ABF20}. What had been \emph{overlooked} in that reference, as well as in other references listed in that reference, is that looking at the \emph{zeros of the branches $f^{\text{M}}_b(p)$}---rather than at the zeros of the original, first-meaning dispersion relation---is \emph{not enough} to determine all the doublers in the case of a \emph{discrete}-time model, such as the QW model. Rather, the definitive way of finding all fermion-doubling problems of a model $\text{M}$ \emph{on its energy branches} is, whether we are in continuous or discrete time, to look at \emph{the symmetries of the $f^{\text{M}}_b(p)$'s}, and more precisely whether the behavior of some $f^{\text{M}}_b(p)$ close to $(E,p)=0$, or, in other words, the shape of its representative curve close to that point, can be found elsewhere in the $(p,E)$-plane in which one plots the $f^{\text{M}}_b(p)$'s. And we see that, to this ``question'', not only does $\text{M}=\text{LGT}$ answer positively, but also $\text{M}=\text{QW}$, since for example on the abovementioned figure the curve from $p=-\pi/\epsilon$ to $0$ is symmetric by a $\pi$ rotation around the Brillouin-zone point $(p,E)=(-\pi/(2\epsilon),\pi/(2\epsilon))$.

Now, finally, in the case of \emph{discrete}-time \emph{naive} fermions of LGT---treated in Ref.\ \cite{DAA2025}---, one would also have, in addition to the spatial doublers already polluting \emph{continuous}-time naive fermions of LGT, extra temporal doublers, but in that case analyzing the FD via the energy branches---as we have just done for the QW thanks to the symmetry arguments---is more complicated, because there is no simple evolution operator that can be defined \emph{a priori}, since it is not a one-time-step scheme but a two-time-step one. This is why we have here limited ourselves to the continuous-time case of naive fermions above.

\stepcounter{footnote}%
\edef\approxfn{\number\value{footnote}}%

\subsection{Comparing between them the spatial-Fourier-transform forms of the
one-time-step GFs of two different lattice models---%
\texorpdfstring{$\mathrm{QW}$}{QW} and (continuous-time)
\texorpdfstring{$\mathrm{LGT}$}{LGT}---, with respect to how well they
\emph{approximate}%
\texorpdfstring{%
  \protect\hyperlink{fn:approximation}{%
    \textsuperscript{\normalfont\approxfn}%
  }%
}{}%
 continuum Dirac fermions}

\footnotetext[\approxfn]{%
\raisedhypertarget{fn:approximation}%
The concept of ``approximation'' we use here is different,
less accurate than that of ``convergence towards'', since these approximations
are sometimes extremely ``far away'' from the (naive) continuum limit, but
still comparable to it. In other words, we cannot always say that the
approximations we will analyze are close to the (naive) continuum limit
\emph{by almost any means}.}

\noindent
{\bfseries Note.} By ``spatial-Fourier-transform forms'', we mean that we have performed the integral over the energy $E$, so that, in  particular, the variable $E$ has disappeared from the mathematical expressions we consider. From now on, hence, whenever the context makes it obvious what the actual meaning of the following is---and so, in particular, in the present subsection---, we will simply say ``Fourier space'' for ``spatial Fourier \mbox{space''--- as} opposed to ``spatiotemporal Fourier space''. \\

\noindent
{\bfseries Analytical formulae.} It is quite trivial to show or simply convince oneself that the one-time-step GF takes the following form for any of the three models $\text{M}= \text{Dirac}, \text{QW}, \text{LGT}$, that is,
\begin{equation}
\label{eq:GenericFourierVX}
G^{\text{M}}(t'+\epsilon,x;t',x') = \int_{-B_{\text{M}}}^{B_{\text{M}}} \frac{dp}{2\pi} e^{ip(x-x')} U^{\text{M}}_{\epsilon}(p)\, .
\end{equation}
In this equation, the bounds of the integral are characterized by $B_\mathrm{M} \equiv p_\mathrm{M}^{\mathrm{max}}$ being  $B_{\text{M}} = +\infty$ for $\text{M}= \text{Dirac}$, and by $B_{\text{M}} = \pi/\epsilon$ for the two other models. \mbox{Moreover, $U^{\text{M}}_\epsilon(p)$ is} the Fourier-space version of the one-step evolution operator $\hat{U}_\epsilon^{\text{M}} \equiv {U}_\epsilon^{\text{M}}(\hat{p})$, or, in other words, it's matrix-valued eigenvalue associated to $\ket p$. This form can be obtained in two different manners---this has been shown explicitly above in the case of the $\text{QW}$, but it can be generalized to the two others models. Indeed, this can be done either by inserting a momentum closure relation in the usual quantum-mechanical propagator, or by performing the integral over the energy variable $E$ in the generic GF form of Eq.\ \eqref{eq:genericG}.

The main thing we want to see is how ``close'' the two discrete models are to their Dirac (naive) continuum limit, that is, how well the \emph{approximate} it, even in situations where one cannot really say, via notions of mathematical convergence, that one is any close to the (naive) continuum limit. This situation interests us because experimentally, in quantum simulation, the continuum limit is very difficult to reach. Of course, ideally, from a naive point of view at least, one would like to compare the one-time-step GFs as functions \mbox{of $\Delta \defeq x-x'$. But,} this is not something that simple analytically \mbox{speaking---as} opposed to numerically---, because in the case of the models $\text{M}= \text{Dirac}$ and $\text{M}= \text{LGT}$, there is no closed form for the above Fourier integral, Eq.\ \eqref{eq:GenericFourierVX}. In the case of $\text{M} = \text{QW}$, remember that our closed form is Eq.\ \eqref{eq:thefinaleqisthisone}. Hence, we will rather compare the integrands of Eq.\ \eqref{eq:GenericFourierVX}, and of course make some comments about what happens when they are integrated.

Now, the QW one-time-step-GF closed form we just mentioned actually restricts the cases on which one should particularly focus while carrying out the comparison we have in mind. For $G^{\text{M}}_{11}$, we will make a special focus on the spatial distance $\Delta = \epsilon$. For $G^{\text{M}}_{12} = G^{\text{M}}_{21}$, the special focus will be on $\Delta = 0$, and, finally, we will not analyze $G^{\text{M}}_{22}$ since it is ``structurally essentially the same'' as $G^{\text{M}}_{11}$. This implies that, if we write Eq.\ \eqref{eq:GenericFourierVX} as
\begin{equation}
\label{eq:GenericFourierVXX}
G^{\text{M}}(t'+\epsilon,x;t',x') = \int_{-B_{\text{M}}}^{B_{\text{M}}} dp \, I^{\text{M}}(p;\epsilon,\Delta) \, ,
\end{equation}
with 
\begin{equation}
I^{\text{M}}(p;\epsilon,\Delta) \defeq \frac{1}{2\pi} e^{ip\Delta} U^{\text{M}}_{\epsilon}(p) \, ,
\end{equation}
the previous announced focuses will be on $I_{11}^{\text{M}}(p;\epsilon,\Delta=\epsilon)$ and $I_{12}^{\text{M}}(p;\epsilon,\Delta=0) = I_{21}^{\text{M}}(p;\epsilon,\Delta=0)$.

But before definitively turning to the aforementioned analyzes, let us give a more explicit and enlightening form for $U^{\text{M}}_{\epsilon}(p)$ and hence $I^{\text{M}}(p;\epsilon,\Delta)$. One can show that, whatever $\text{M}= \text{Dirac}, \text{LGT}, \text{QW}$, we have
\par\nobreak
{\footnotesize
\begin{equation}
\label{eq:theIs}
I^{\text{M}}(p;\epsilon,\Delta) = \frac{1}{2 \pi}e^{ip\Delta} 
\begin{bmatrix}
a^{\text{M}}(p;\epsilon) - i b^{\text{M}}(p;\epsilon) & - i c^{\text{M}}(p;\epsilon) \\
- i c^{\text{M}}(p;\epsilon) & a^{\text{M}}(p;\epsilon) + i b^{\text{M}}(p;\epsilon)
\end{bmatrix} \, ,
\end{equation}}
where the expressions $e^{\text{M}}(p;\epsilon)$, $e = a, b, c$, are real-valued. These expressions are given in App.\ \ref{app:theintegrands} for each of the three models.

There is still one thing we ought to do before analyzing plots of the expressions, namely, reducing the number of variables to the bare minimum. In order to do so, we think one should start by the following remark. The reader may have noticed one particular difference between the continuum model and its two different discretizations, which is going to be of particular importance in the analysis of the Green's functions: the bounds of the momentum integral for the continuum model \mbox{are $B_{\text{Dirac}}=\pm \infty$, but} $B_{\text{M}}=\pm \frac{\pi}{\epsilon}$, respectively, for its two discretizations $\text{M} = \text{QW}$ and $\text{M} = \text{LGT}$. This poses a difficulty: at some point in the analysis, we will for sure be interested in the (naive) continuum limit, usually expressed, in its simplest formal way, as $\epsilon \rightarrow 0$; but as we do so the bounds of the discrete model vary, they increase up to $\pm\infty$, and this seems very unpractical to us. Moreover, there is actually a very natural change of variable suggested by the various expressions of Eq.\ \eqref{eq:theIs}---given in App.\ \ref{app:theintegrands}. This change of variable is $p \rightarrow P \defeq p \epsilon$, which will make the bounds of the discrete models become $\epsilon$-independent numerical values, namely, $B'_{\text{M}} = \pm \pi$. Notice that the bounds of the continuum model will still be $\pm \infty$, and this difference with respect to the discrete models must still be discussed, but at least now the bounds of the discrete models are fixed, i.e., $\epsilon$-independent.

After performing the announced change of variable, the Fourier expression of the Green's function in Eq.\ \eqref{eq:GenericFourierVXX} becomes
\begin{equation}
\label{eq:GenericFourier2}
G^{\text{M}}(t'+\epsilon,x;t',x') = \frac{1}{\epsilon} \int_{-B'_{\text{M}}}^{B'_{\text{M}}} dP J^{\text{M}}(P;X) \, ,
\end{equation}
where $X\defeq (x-x')/\epsilon$, and
\begin{equation}
J^{\text{M}}(P;X) \defeq \frac{1}{2\pi} e^{iPX} V^{\text{M}}(P) \, ,
\end{equation}
with $V^{\text{M}}(P)$ being the expression $U^{\text{M}}_{\epsilon}(p)$ but written with a (matrix-valued) function of $P$. Notice that, contrary to $U^{\text{M}}_{\epsilon}(p)$, we have not written
any $\epsilon$ dependence on $V^{\text{M}}(P)$: this is because, after performing the change of variable $p \rightarrow P$, and after extracting the $1/\epsilon$ factor visible in \mbox{Eq.\ \eqref{eq:GenericFourier2}}, in $V^{\text{M}}(P)$ there is no more $\epsilon$ dependence \emph{except via the expression} 
\begin{equation}
M \equiv \mu(m,\epsilon) \defeq m \epsilon \, ,   
\end{equation}
---where we recall that $m$ is the mass of the Dirac field of the (naive) continuum limit---, and it is this whole \mbox{product $M= \epsilon m$} that we will be varying rather than just $\epsilon$. This is more than something anecdotical, it is a profound aspect of taking the (naive) continuum limit: ``taking both $p \epsilon $ going to zero on the one hand, and $m\epsilon$ going to zero on the other hand'' is a non-equivalent and much more enlightening way of taking the (naive) continuum limit than $\epsilon \rightarrow 0$, a way which is far less formal, more physical, and which demands a more subtle analysis. Going beyond the continuum-limit perspective, we will actually precisely use this ``double freedom'', in $P$ and $M$, respectively, rather than the ``single freedom'' in $\epsilon$, in order to explore regimes where only $P$or only $M$ is small, so that one cannot really say one is any close to the continuum limit\footnote{Actually, even taking $P$ AND $M$ small is not equivalent to taking $\epsilon$ small, and so does not guarantee we are any close to the continuum limit: the right, i.e., the physical way of taking the continuum limit, is taking both $P$ and $M$ small, but to discuss the fact that, in $P=p\epsilon$ and $M=m\epsilon$, although the smallness of $\epsilon$ is indeed controlled jointly, at once, for both parameters, $p$ can be extremely small or extremely large, and same for $m$, which does have an influence on a truly physical continuum limit---rahter than a purely formal one.}, but one can still compare the regime under study to that continuum limit---nothing forbids us to doing it. Also, we have extracted the $1/\epsilon$ because it is not relevant. In the case of the QW, you can see this $1/\epsilon$ in the final closed form of Eq.\ \eqref{eq:thefinaleqisthisone}: if we stay in the discrete setting, it will cancel with the factor $\epsilon$ of the propagation equation, Eq.\ \eqref{eq:propagationequation}; and if we go to the continuum, it will combine with (i.e., multiply to) the Kronocker deltas of the final closed form to produce Dirac delta \mbox{functions---while} the factor $\epsilon$ of the propagation equation ``yields'' the integration element $dx$.

One extra particular choice must be made and commented. Since the discrete models are both at least on a spatial lattice, and with spacing \emph{imposed} by the QW model as being equal to the time step $\epsilon$\footnote{What matters is the fact that the scaling with $\epsilon$ is linear, since any constant multiplicative factor will just represent a choice of spatial units. Since we have taken the speed of light $c=1$, this factor is just $1$ in our case.}---``imposed'' in the sense that we must choose the same for the other models if we want to be able to compare sensefully all models---, the cases where $x-x'$, in the continuum model, is not a (relative) multiple of $\epsilon$, do not interest us. We will call $D \in \mathbb Z$ this multiple, that is,
\begin{equation}
\label{eq:defD}
D \defeq \frac{x-x'}{\epsilon} \, .    
\end{equation}
This implies that in the end, Eq.\ \eqref{eq:GenericFourier2} becomes
\begin{equation}
\label{eq:GenericFourier3}
G^{\text{M}}(t'+\epsilon,x;t',x') = \frac{1}{\epsilon} \int_{-B'_{\text{M}}}^{B'_{\text{M}}} dP \, F^{\text{M}}(P;D) \, ,
\end{equation}
with 
\begin{equation}
F^{\text{M}}(P;D) \defeq \frac{1}{2\pi} e^{iPD} V^{\text{M}}(P) \, ,   
\end{equation}
which, remember, is a $2\times 2$ complex matrix.

Finally, let us write the sole expressions we will in the end be plotting. For the Dirac, continuum model, these expressions are
\par\nobreak
{\small
\begin{subequations}
\begin{align}
\mathrm{Re}\{ e^{-iPD}F^{\text{Dirac}}_{11}(P;D) \} &= \frac{1}{2\pi}  A^{\text{Dirac}}(P;M) \\
&= \frac{1}{2\pi}   \cos(\sqrt{P^2+M^2}) \label{eq:ADiracPM}\\
- \mathrm{Im}\{ e^{-iPD} F^{\text{Dirac}}_{11}(P;D) \} &= \frac{1}{2\pi}  B^{\text{Dirac}}(P;M) \\
&= \frac{1}{2\pi} P \, \mathrm{sinc}(\sqrt{P^2+M^2}) \\
- \mathrm{Im}\{ e^{-iPD} F^{\text{Dirac}}_{i,j\neq i}(P;D) \} &= \frac{1}{2\pi}  C^{\text{Dirac}}(P;M) \\
&= \frac{1}{2\pi} M \, \mathrm{sinc}(\sqrt{P^2+M^2}) \, .
\end{align}    
\end{subequations}}
For the LGT model, these expressions are
\par\nobreak
{\small
\begin{subequations}
\begin{align}
\mathrm{Re}\{e^{-iPD} F^{\text{LGT}}_{11}(P;D) \} &= \frac{1}{2\pi}  A^{\text{LGT}}(P;M) \\
&= \frac{1}{2\pi} \cos\left(\sqrt{\sin^2(P)+M^2}\right) \\
- \mathrm{Im}\{ e^{-iPD} F^{\text{LGT}}_{11}(P;D) \} &= \frac{1}{2\pi}  B^{\text{LGT}}(P;M) \\
&= \frac{1}{2\pi} \sin(P) \, \mathrm{sinc}\left(\sqrt{\sin^2(P)+M^2}\right)  \\
- \mathrm{Im}\{ e^{-iPD} F^{\text{LGT}}_{i,j\neq i}(P;D) \} &= \frac{1}{2\pi}  C^{\text{LGT}}(P;M) \\
&= M \, \mathrm{sinc}\left(\sqrt{\sin^2(P)+M^2} \right) \, .
\end{align}    
\end{subequations}}
For the QW model, these expressions are
\par\nobreak
{\small
\begin{subequations}
\begin{align}
\mathrm{Re}\{ e^{-iPD} F^{\text{QW}}_{11}(P;D) \} &= \frac{1}{2\pi}  A^{\text{QW}}(P;M) \\
&= \frac{1}{2\pi} \cos(M)\cos(P)  \\
- \mathrm{Im}\{ e^{-iPD} F^{\text{QW}}_{11}(P;D) \} &= \frac{1}{2\pi}  B^{\text{QW}}(P;M) \\
&= \frac{e^{iPD}}{2\pi} \cos(M)\sin(P)  \\
- \mathrm{Im}\{ e^{-iPD} F^{\text{QW}}_{i,j\neq i}(P;D) \} &= \frac{1}{2\pi}  C^{\text{QW}}(P;M) \\
&= \frac{1}{2\pi} \sin(M) \, .
\end{align}    
\end{subequations}}
Notice the \emph{extreme} simplicity of the QW case. In this QW case, we have (we knew this from the ``very beginning''), for the $11$ and $22$ coefficients, a complex-exponential form for the expression in factor of $e^{iPD}/(2\pi)$---which simplifies computations and understanding in certain situations---, but not for the two other cases---which however of course come from a \emph{matrix} exponential. 

Remember that the sole remaining $\epsilon$ dependence is \mbox{via $P\defeq p\epsilon$} and $M\defeq m\epsilon$, which will be our privileged variables for the study.

Finally, regarding the \emph{discrete}-time version of the real-time naive-LGT free-Dirac-equation discretization, which we do not treat here---but quite some technical information is given about it in Ref.\ \cite{DAA2025}---, one can show that it cannot be obtained from a one-time-step discrete scheme with \emph{unitary} evolution operator (unpublished yet)---but a priori this does not imply that one cannot find a non-unitary such evolution operator (which may not be unique) that takes $\psi(t)$ to $\psi(t+\epsilon)$. \\

\begin{figure*}
    \centering
    \includegraphics[width=0.8\linewidth]{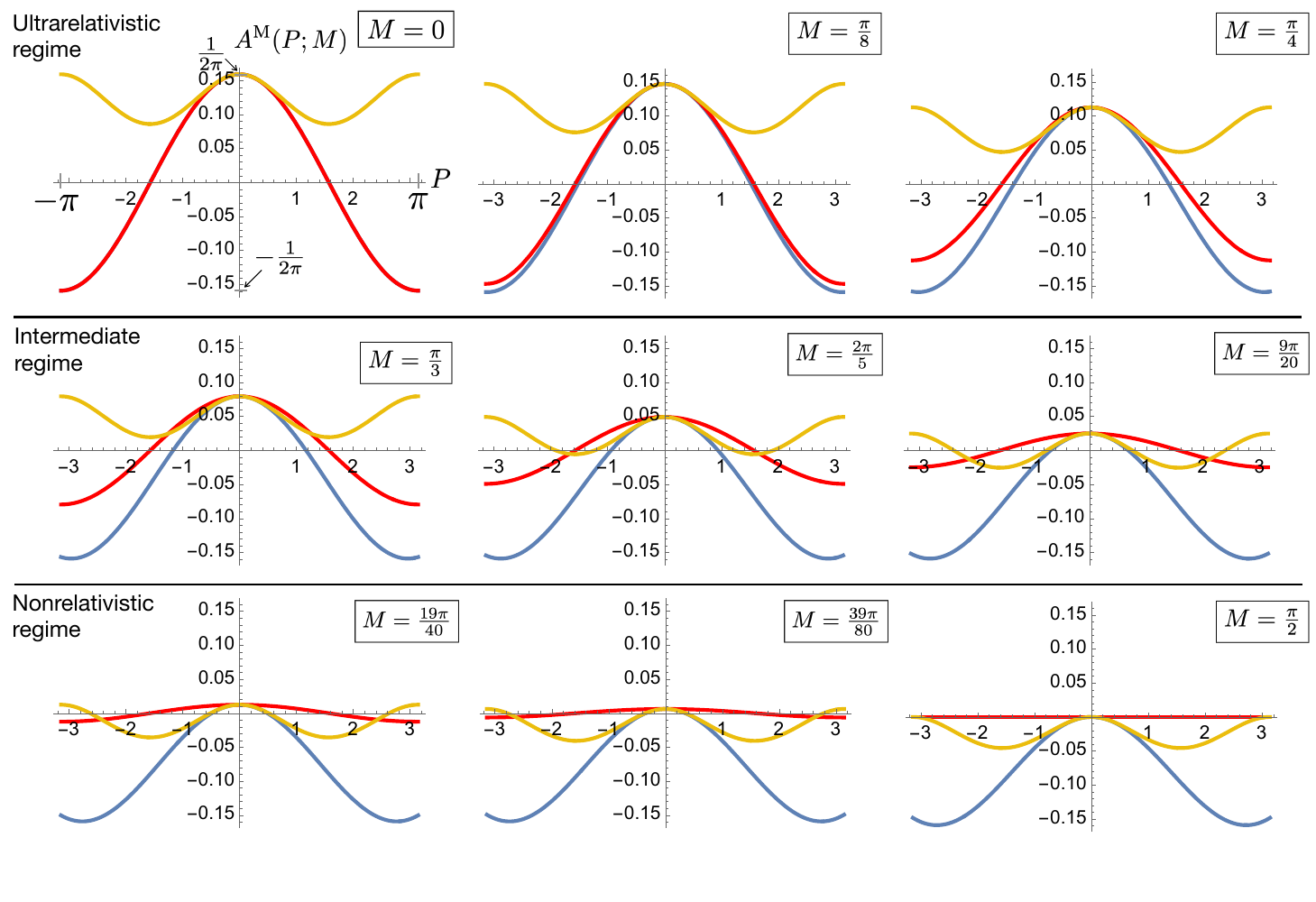}
    \caption{Comparison between the $A^{\text{M}}(P;M)/(2\pi)$'s for the three models $\text{M} = \text{Dirac}$ (blue), $\text{M} = \text{QW}$ (red), and $\text{M} = \text{LGT}$ (gold), as functions of $P\defeq \epsilon p$, for 9 values of $M\defeq \epsilon m$. Remember that these $A^{\text{M}}(P;M)/(2\pi)$'s are the factors, within the integrands of the momentum integral yielding the one-time-step Green's function, which are independent from the considered transition distance $D = (x-x')/\epsilon \in \mathbb Z$. We have called the top row ``Ultrarelativistic regime'' because of the following reason: in this regime, the rescaled mass $M$ is small enough with respect to, say, the maximum $P=\pi$ that the rescaled momentum $P$ takes for the discrete models---actually, because of symmetry reasons, the reference value should rather be $\pi/2$---, in such a way that the QW red curve is overall much closer to its naive continuum limit than the LGT gold curve. In other words, the QW model is, on the basis of the present figure at least, a much better approximation of the ``naive continuum model'' (i.e., ``reached'' via a \emph{naive} continuum limit) in such an ultrarelativistic regime. We have called the middle row ``Intermediate regime'' because of the following reason: in this regime, $M$ is not small enough with respect to $\pi$, but not close enough to $M=\pi/2$. In this regime, it becomes less obvious to say whether it is the QW model or the LGT model that is closer to the naive continuum model. We have called the bottom row ``Nonrelativistic regime'' because of the following reason: in this regime, the rescaled mass $M$ is close enough to $\pi/2$, in such a way that the LGT gold curve is overall closer to its naive continuum limit than the QW red curve. In other words, the LGT model is, on the basis of the present figure at least, a better approximation of the naive continuum model in such a nonrelativistic regime.}
    \label{fig:As}
\end{figure*}

\noindent
{\bfseries Study of the $D$-independent part of the integrands.} We are here going to study the expressions $(1/(2\pi))E^{\text{M}}_F(P;M)$, $E=A,B,C$, as functions of $P$, for different values of $M$. In Fig.\ \ref{fig:As}, we show plots comparing, each, the three $A^{\text{M}}(P;M)$'s as functions of $P$, each plot being for a different value of $M$.

Before commenting on the shapes of the curves of these plots, let us first notice that the continuum-model case has been plotted over the same bounded \mbox{interval $[-\pi,\pi]$ as} that of the integral of the discrete models, whereas the integral of the continuum model is actually over $]-\infty,+\infty[$. This is because it is easy to convince oneself either numerically---by plotting all continuum-model plots of Fig.\ \ref{fig:As} over an \mbox{interval $[-N\pi,N\pi]$ with} $N$ large enough (a few units is actually enough)---, or analytically by doing a limit of large $P$ in Eq.\ \eqref{eq:ADiracPM}, that $A^{\text{Dirac}}(P;M)$ is \emph{periodically oscillating around $0$} when $|P|\gg M$, so that, after multiplying $A^{\text{Dirac}}(P;M)$ by $e^{iPD}$, we still have a function (complex-valued this time) whose real and imaginary parts periodically oscillate around $0$ for $|P|\gg M$, and this implies that the contribution, to the momentum integral over $]-\infty,+\infty[$, of the region $]-\infty,-\pi] \cup [\pi, +\infty[$, is negligible.

Let us now give extra comments on Fig.\ \ref{fig:As} with respect to what is already said in its caption. For a vanishing mass $M=0$, the QW simply describes its naive continuum limit \emph{exactly}---contrary to the LGT case---, which is why we cannot see the blue curve, and this whatever $\epsilon$, although, of course, the difference between both models is that if $\epsilon\neq 0$ the QW is only defined on the spacetime lattice of spacing $\epsilon$, whereas its naive continuum limit, the $(1+1)$D Dirac equation, is defined over the whole $(1+1)$D spacetime continuum. For the first relevant maximal rescaled mass $M = \pi/2$ (the cases with higher masses can be obtained by symmetries), the studied integrand vanishes for the QW model, and the Dirac-equation naive continuum limit of the QW model cannot be obtained at all. Still for $M=\pi/2$, the LGT still approaches correctly the continuum-limit curve for small-enough $P$---as for any value of $M$ actually. So, as a qualitative conclusion: the QW model \emph{approximates}\footnote{Again, we talk about an \emph{approximation} here, not about a \emph{properly defined convergence} towards the naive continuum limit, since we \emph{already know} that \emph{both} discrete models can be made to converge towards this naive continuum limit for $M \rightarrow 0$, provided $P\rightarrow 0$.} the naive continuum limit \emph{better} than the LGT model for small $M$'s, and it is the contrary for $M$ close enough to $\pi/2$. This qualitative conclusion is not so surprising \emph{naively}, since the continuous-time LGT model is a model that does not have a \emph{strict} lightcone, since one can move from a strictly finite distance $\epsilon$ in an infinitesimal amount of time $dt$, which can of course be faster than the speed of light.

For the sake of completeness, we provide in App.\ \ref{app:otherfigures} the same type of figure as Fig.\ \ref{fig:As} but for $B^{\text{M}}(P;M)$ on the one hand (Fig.\ \ref{fig:Bs}), and for $C^{\text{M}}(P;M)$ on the other \mbox{hand (Fig.\ \ref{fig:Cs})}. \\

\noindent
{\bfseries Study of the integrands for relevant values of $D$.} We are now going to study the following expressions,
\par\nobreak
{\footnotesize
\begin{subequations}
\begin{align}
R_{11}^{\text{M}}(P;M,D=1) &\defeq \text{Re}\{F_{11}^{\text{M}}(P;M,D=1)\} \\
&= \frac{1}{2 \pi} \left[ \cos(P) A^{\text{M}}(P;M) + \sin(P) B^{\text{M}}(P;M)\right] \\
Y_{11}^{\text{M}}(P;M,D=1) &\defeq \text{Im}\{F_{11}^{\text{M}}(P;M,D=1)\} \\
&= \frac{1}{2 \pi} \left[ \sin(P) A^{\text{M}}(P;M) - \cos(P) B^{\text{M}}(P;M)\right] \, ,
\end{align}
\end{subequations}}
and this will actually be enough, since a few lines of computation show that
\par\nobreak
{\small
\begin{subequations}
\begin{align}
R_{i,j\neq i}^{\text{M}}(P;M,D=0) &\defeq \text{Re}\{F_{i,j\neq i}^{\text{M}}(P;M,D=0)\} \\
&= 0 \\
Y_{i,j\neq i}^{\text{M}}(P;M,D=0) &\defeq \text{Im}\{F_{i,j\neq i}^{\text{M}}(P;M,D=0)\} \\
&= -\frac{1}{2 \pi} C^{\text{M}}(P;M) \, ,
\end{align}
\end{subequations}}
and the last expression has already been plotted in \mbox{App.\ \ref{app:otherfigures}}.

\begin{figure*}
    \centering
    \includegraphics[width=0.8\linewidth]{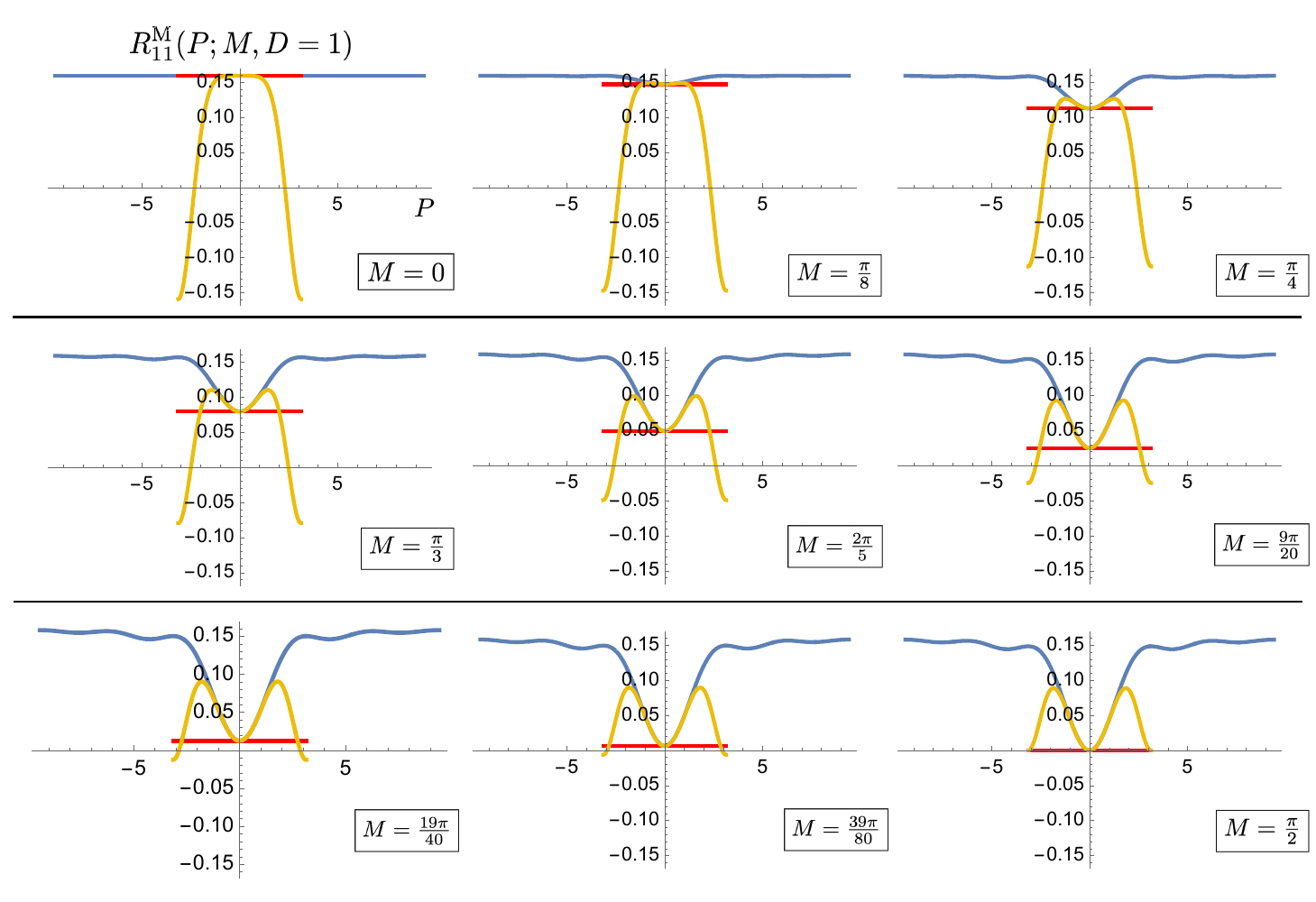}
    \caption{Comparison of the real parts $R_{11}^{\text{M}}(P;M,D=1)$ of the $11$ matrix components of the Fourier integrands of the one-time-step Green's functions of the three models $\text{M} = \text{Dirac}$ (blue), $\text{M} = \text{QW}$ (red), and $\text{M} = \text{LGT}$ (gold), respectively, for a transition of $D=+1$ spatial lattice step (in the direction of growing positions, hence the $+$ in $D=+1$). The lecture order of the plots in the order of increasing rescaled masses $M$ is, graphically, $\doublezigzag$, from the most relativistic situation for $M=0$ in the top left corner---``relativistic'' in the sense defined in Fig.\ \ref{fig:As}---, to the less relativistic one for $M=\pi/2$ in the bottom right corner. We see that, the more the situation is relativistic (i.e., the smaller the rescaled mass is), the more the Dirac Fourier integrand is just a constant over the whole integration interval $\mathbb R$, which explains why ``the result of the integral is $+\infty$'', as it should for the $u=u^{\prime}$ ``value'' of a Dirac delta function $\delta(u-u^{\prime})$, where here $u=x-\epsilon$ and $u^{\prime}=x$---actually, the situation is a bit more complicated, see App.\ \ref{app:TaylorExpansion}.  For $\text{M} = \text{QW}$, the situation is similar, but the value of the integral is, this time, finite, and equal to $1$ for $m=0$ (and hence $M=0$), as it should for the $u=u^{\prime}$ value of a Kronecker symbol $\delta_{u,u^{\prime}}$, and as one can explicitly see in Eq.\ \eqref{eq:thefinaleqisthisone}, for the term $11$. \label{fig:R11}}
\end{figure*}

\begin{figure*}
    \centering
    \includegraphics[width=0.8\linewidth]{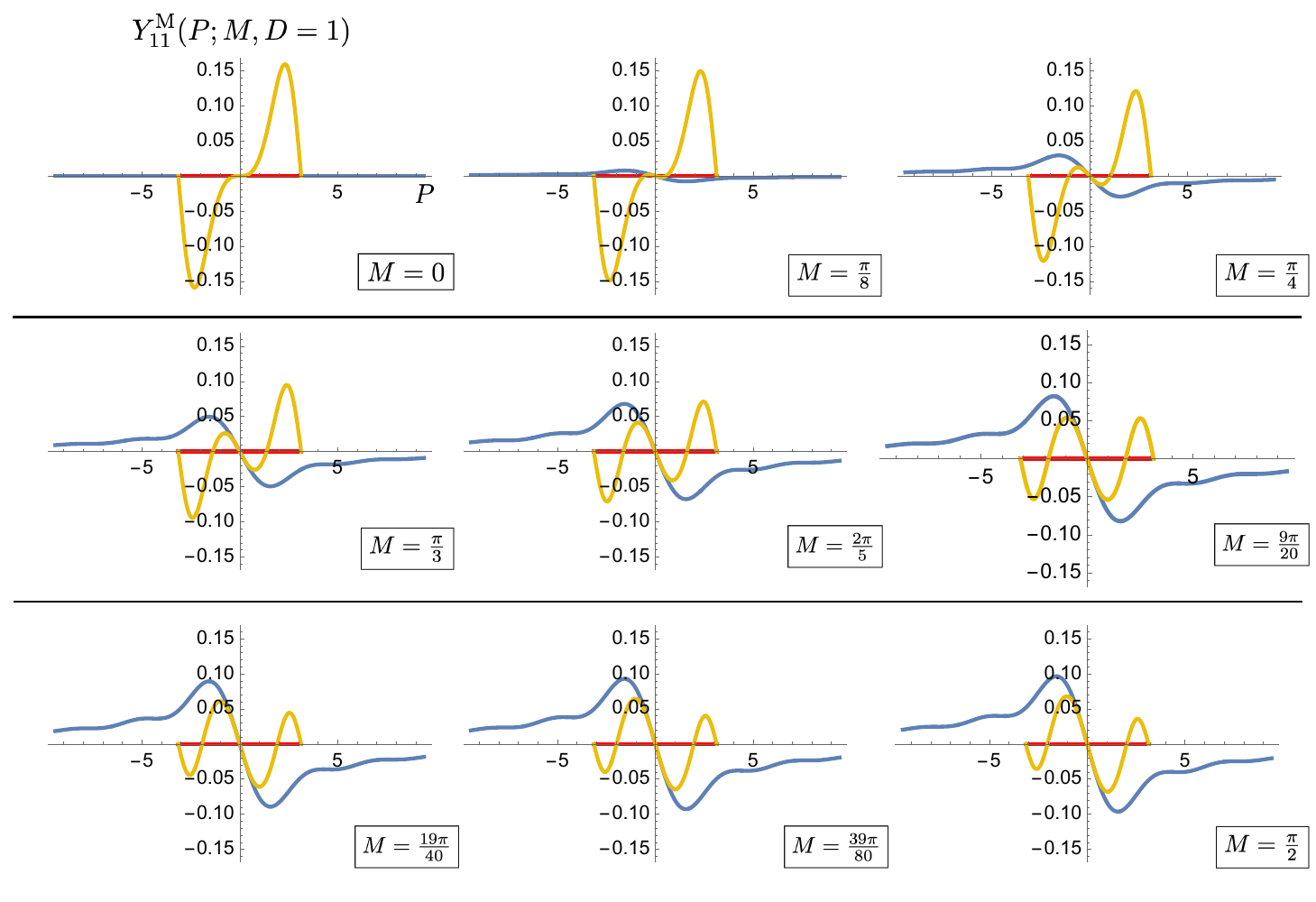}
    \caption{Same as Fig.\ \ref{fig:R11} but for the imaginary parts parts $Y_{11}^{\text{M}}(P;M,D=1)$---of the $11$ matrix components of the Fourier integrands of the one-time-step Green's functions of the three models $\text{M} = \text{Dirac}$ (blue), $\text{M} = \text{QW}$ (red), and $\text{M} = \text{LGT}$ (gold), respectively. This imaginary part is zero for the QW model. It is also zero for the naive continuum model in the case $M=0$. The more $M$ gets close to $\pi/2$ coming from lower values (starting at $0$) and increasing, the more the LGT model approximates well the naive continuum model, although of course this is just an approximation, which is the best when $P=p\epsilon$ is small enough, that is, close to $P=0$, but this is not an exact match of the naive continuum limit since $M$ gets close to a non-zero value ($\pi/2$).}
    \label{fig:I11}
\end{figure*}

The plots are shown in Figs.\ \ref{fig:R11} and \ref{fig:I11}, and the comments are made in the captions. \\

\section{Correlation function of the Flavored Dirac QCA (FDQCA), which is the FD-fixed model}
\label{sec:correlationFQCA}

\subsection{Computation of the FDQCA Green's function}

\subsubsection{Recap on the Flavored Dirac QCA}

\noindent
In Ref.\ \cite{DAA2025}, we have proposed a solution to solve the FD problem of the Dirac QCA, problem which has been reexplained in detail in Sec.\ \ref{subsec:witnessingFD}. We have called this solution ``flavor staggering only'', because it is based on the staggering of extra flavor degrees of freedom that we introduced, but without staggering the chirality. This solution amounts to replacing, in the continuum, a theory of one FD-suffering Dirac fermion, by a theory of two completely independent FD-free Dirac fermions, each one carrying a flavor. This independence means that the flavor degree of freedom is, in particular, not entangled with the dynamics---and not purely classically correlated to it either---, at least in the continuum theories we have just mentioned. Regarding our ``flavor staggering only'' model, which we call Flavored Dirac QCA (FDQCA), and which is of course in \emph{discrete} spacetime, things are a bit more complicated regarding the entanglement between the flavor degree of freedom and the chirality, but we will not enter into any details here. What matters here for us is that---and this is one of main aspects of our solution to have in mind---the BZ has now a rhombus shape, see Fig.\ 3 of \mbox{Ref.\ \cite{DAA2025}}.

For the sake of completeness, we recall the form that the wavefunction of the Dirac FQW must have if we want to be able to reach the continuum limit:
\begin{equation}
    \psi^{\text{FQW}}(t,x) = \chi(t,x) \otimes \sigma_x^{(t+x)/\epsilon}\ket 0 \, ,
\end{equation}
where $\sigma_x$ is the first Pauli matrix, $\ket 0 \equiv (1,0)^{\top}$ is the first basis vector of the flavor Hilbert space, and $\top$ is the transposition operation. Also, $\chi(t,x)$ belongs to the Hilbert space of the (non-flavored) QW, but it contains the two flavored wavefunctions: when $(t+x)/\epsilon$ is even, then $\chi(t,x)=\psi_0(t,x)$, the wavefunction of the first flavor, and  when $(t+x)/\epsilon$ is odd, then $\chi(t,x)=\psi_1(t,x)$, the wavefunction of the second flavor. All this being said, the knowledge of the previous equation is not necessary for what follows, which focuses on the Green's function of the model, which is independent of any initial condition and subsequent form that the wavefunction may have at an arbitrary time step $n$.

\subsubsection{Looking for an appropriate GF}

\noindent
This rhombus shape we have for the BZ implies in particular that a suitable ansatz form with which to look for the Green's function of the FDQCA is
\par\nobreak
{\small
\begin{align}\label{eq:propagatorFQCA}
	&G^{\text{FQW}}(t,x ; t',x') \nonumber \\
&=  2 \int_{- \pi/\epsilon}^{\pi/\epsilon}  \frac{dp}{2 \pi }   e^{ip(x-x')} \int_{- \pi/\epsilon+|p|}^{\pi/\epsilon- |p|} \frac{dE}{2 \pi }   \hspace{1mm}  
			e^{-iE(t-t')}  G^{\text{FQW}}_{\mathcal{B_{\text{FQW}}}}(E,p) \, ,
    \end{align}}
instead of the ansatz of Eq.\ \eqref{eq:propagator}. The factor of $2= (\sqrt 2)^2 $ in front of the RHS is necessary for a correct normalization.

The GF definition equation for the FQW reads
\begin{equation}
\label{eq:defeqGFQW}
\mathcal{L}^{\text{FQW}}_{\epsilon} G^{\text{FQW}}(\cdot,\cdot;t',x') \big|_{t,x} = i \left(\frac{1}{\epsilon}\delta_{t,t'}\right) \left(\frac{1}{\epsilon}\delta_{x,x'}\right)  \, ,
\end{equation}
with 
\begin{equation}
\mathcal{L}^{\text{FQW}}_{\epsilon} \defeq   \frac{1}{i \epsilon} (\mathcal T_\epsilon^{f} \, \mathcal U_\epsilon^{\text{FQW}} - \mathds 1^{\text{FQW}}) \, , 
\end{equation}
where (i)
\begin{equation}
\mathcal T_\epsilon^{f} \defeq \mathcal T_\epsilon \otimes \sigma_x \, ,
\end{equation}
with $\sigma_x$ acting on the flavor space, and where (ii) 
\par\nobreak
{\small
\begin{equation}
\mathcal U_\epsilon^{\text{FQW}} \defeq \begin{bmatrix}
				\cos(m \epsilon) \mathcal{S}_{\epsilon} \otimes \sigma_x &  - i \sin(m \epsilon) \otimes \mathbb I_{\text{flav.}} \\
				- i \sin(m \epsilon)\otimes \mathbb  I_{\text{flav.}} & \cos(m \epsilon) \mathcal{S}^{\dagger}_{\epsilon} \otimes \sigma_x
			\end{bmatrix} \, ,      
\end{equation}}
is the position-representation version of Eq.\ (4.7c) of \mbox{Ref.\ \cite{DAA2025}}.

In Fourier space, Eq.\ \eqref{eq:defeqGFQW} becomes
\begin{equation}
\label{eq:defeqinFS}
{L}^{\text{FQW}}_{\epsilon}(E,p)G^{\text{FQW}}(E,p) = i \,  \mathbb I_{\text{int.}} \, ,
\end{equation}
where $\mathbb I_{\text{int.}} \defeq \mathbb I_{\text{chir.}} \otimes \mathbb I_{\text{flav.}}$, and where, for the sake of completeness, we give explicitly
\par\nobreak
{\small
\begin{align}
\label{eq:final_expression_bis}
{L}^{\text{FQW}}_{\epsilon}(E,p)&=  \frac{1}{i \epsilon}
\begin{bmatrix}
\big( c_\epsilon e^{i \epsilon E} e^{-i\epsilon p} - 1 \big) \, \mathbb I_{\text{flav.}} &  i s_\epsilon e^{i \epsilon E} \, \sigma_x  \\
 i s_\epsilon e^{i \epsilon E} \,  \sigma_x   & \big( c_\epsilon e^{i \epsilon E} e^{i\epsilon p} - 1  \big) \, \mathbb  I_{\text{flav.}}
\end{bmatrix}  \, .
\end{align}}

Now, it is trivial to show that a solution $G^{\text{FQW}}(E,p)$ of Eq.\ \eqref{eq:defeqinFS} is
\begin{equation}
G^{\text{FQW}}(E,p) = 
\begin{bmatrix}
G^{\text{QW}}_{11}(E,p) \, \mathbb  I_{\text{flav.}}& G^{\text{QW}}_{12}(E,p) \, \sigma_x   \\
 G^{\text{QW}}_{21}(E,p)  \,\sigma_x   & G^{\text{QW}}_{22}(E,p) \, \mathbb  I_{\text{flav.}} 
\end{bmatrix}    \, ,
\end{equation}
where we recall that $G^{\text{QW}}(E,p)$ is given by Eq.\ \eqref{eq:final_expression_bis0}, and satisfies the Fourier-space version of Eq.\ \eqref{eq:thedefeq}.

\subsubsection{Finding the simplest expression for the direct-space FQW GF}
\label{subsubsec:findingthesimplestexpression}

\noindent
The difficulty here is the new, rhombus BZ. We will first perform some changes of variables that will make us able to compute the integral of Eq.\ \eqref{eq:propagatorFQCA} in half the situations we encounter.

We rewrite Eq.\ \eqref{eq:propagatorFQCA} as 
\par\nobreak
{\small
\begin{align}\label{eq:cuttingtheintegral}
	G^{\text{FQW}}(t,x ; t',x')
= I^{(1)}(D,\Delta n) + I^{(2)}(D,\Delta n) + I^{(3)}(D,\Delta n) \, ,
    \end{align}}
with
\begin{subequations}
\label{eqs:integrals0}
\begin{align}
&I^{(1)}(D,\Delta n) \nonumber  \\
&\defeq  2 \int_{- \pi/\epsilon}^{0} \frac{dp}{2 \pi }  \, e^{ipD\epsilon}  \int_{- \pi/\epsilon- p}^{0} \frac{dE}{2 \pi }   \hspace{1mm}  
			e^{-iE\Delta n \e}  G^{\text{FQW}}_{\mathcal{B_{\text{FQW}}}}(E,p) \\
&I^{(2)}(D,\Delta n) \nonumber \\
&\defeq 2 \int_{- \pi/\epsilon}^{0} \, \frac{dp}{2 \pi } \, e^{ipD\epsilon} \int_{0}^{\pi/\epsilon+p} \frac{dE}{2 \pi }   \hspace{1mm}  
			e^{-iE\Delta n \e}  G^{\text{FQW}}_{\mathcal{B_{\text{FQW}}}}(E,p)   \\
&I^{(3)}(D,\Delta n) \nonumber \\
&\defeq 2 \int_{0}^{\pi/\epsilon} \frac{dp}{2 \pi } \, e^{ipD\epsilon}   \int_{-\pi/\epsilon+p}^{\pi/\epsilon-p} \frac{dE}{2 \pi }   \hspace{1mm}  
			e^{-iE\Delta n \e}  G^{\text{FQW}}_{\mathcal{B_{\text{FQW}}}}(E,p) \, ,
\end{align}    
\end{subequations}
with $D \defeq (x-x')/\epsilon$ having already been defined in \mbox{Eq.\ \eqref{eq:defD},} and
\begin{equation}
 \Delta n \defeq \frac{t-t'}{\epsilon}   \, . 
\end{equation}

Let us now do the announced changes of variable, which are for the first and second integrals. For the first integral, we do the following two changes of variable,
\begin{subequations}
\begin{align}
E &\longrightarrow E' = E + \frac{\pi}{\e} \\
p &\longrightarrow p' = p + \frac{\pi}{\e} \, ,
\end{align}    
\end{subequations}
and for the second, the following two,
\begin{subequations}
\begin{align}
E &\longrightarrow E'' = E - \frac{\pi}{\e} \\
p &\longrightarrow p' = p + \frac{\pi}{\e} \, ,
\end{align}    
\end{subequations}
so that in the end Eqs.\ \eqref{eqs:integrals0} become (only the first and the second have been modified), after renaming $(p',E')$ by $(p,E)$ in the first integral, and $(p',E'')$ by $(p,E)$ in the second integral,
{\small
\par
\nobreak
\begin{subequations}
\label{eqs:integrals}
\begin{align}
I^{(1)}(D,\Delta n) 
&=  2 (-1)^{D+\Delta n} \int_{0}^{\tfrac{\pi}{\e}}  \frac{dp}{2 \pi }  \,e^{ipD\epsilon}   \int_{\tfrac{\pi}{\e}-p}^{\tfrac{\pi}{\e}} \frac{dE}{2 \pi }   \hspace{1mm}
			e^{-iE\Delta n \e}  \nonumber  \\
   & \ \ \ \ \ \ \ \ \ \ \ \ \ \ \ \ \ \ \ \ \ \  \times G^{\text{FQW}}_{\mathcal{B_{\text{FQW}}}}(E-\tfrac{\pi}{\e},p-\tfrac{\pi}{\e}) \\
I^{(2)}(D,\Delta n) 
&= 2 (-1)^{D+\Delta n} \int_{0}^{\tfrac{\pi}{\e}} \frac{dp}{2 \pi } \, e^{ip D\epsilon} \int_{-\tfrac{\pi}{\e}}^{-\tfrac{\pi}{\e}+p} \frac{dE}{2 \pi }   \hspace{1mm}  
			e^{-iE\Delta n \e} \nonumber\\ 
& \ \ \ \ \ \ \ \ \ \ \ \ \ \ \ \ \ \ \ \ \ \  \times G^{\text{FQW}}_{\mathcal{B_{\text{FQW}}}}(E+\tfrac{\pi}{\e},p-\tfrac{\pi}{\e})   \\
I^{(3)}(D,\Delta n) 
&= 2 \int_{0}^{\tfrac{\pi}{\e}} \frac{dp}{2 \pi } \, e^{ip D\e}  \int_{-\tfrac{\pi}{\e}+p}^{\tfrac{\pi}{\e}-p} \frac{dE}{2 \pi }   \hspace{1mm}  
			e^{-iE\Delta n \e}   \nonumber \\
& \ \ \ \ \ \ \ \ \ \ \ \ \ \ \ \ \ \ \ \ \ \  \times G^{\text{FQW}}_{\mathcal B_{\text{FQW}}}(E,p) \, .
\end{align}    
\end{subequations}}
Also, it is trivial to show that
\begin{equation}
G^{\text{FQW}}_{\mathcal{B_{\text{FQW}}}}(E\pm\frac{\pi}{\e},p\pm\frac{\pi}{\e})  = 
G_{\text{BZ};\text{d.}}^{\text{FQW}}(E,p) - G_{\text{BZ};\text{n.d.}}^{\text{FQW}}(E,p)    \, ,
\end{equation}
where we have used the notation $\text{BZ} \defeq \mathcal B_{\text{FQW}}$, and where
{\small
\par
\nobreak
\begin{subequations}
\begin{align}
G_{\text{BZ};\text{d.}}^{\text{FQW}}(E,p) &\defeq
\begin{bmatrix}
G_{\mathcal{B};11}^{\text{QW}}(E,p) \, \mathbb I_{\text{flav.}}& 0 \\
0 & G_{\mathcal{B};22}^{\text{QW}}(E,p) \, \mathbb I_{\text{flav.}}
\end{bmatrix}  \, , \label{eq:diagG}\\
G_{\text{BZ};\text{n.d.}}^{\text{FQW}}(E,p) & \defeq
\begin{bmatrix}
0 & G_{\mathcal{B};12}^{\text{QW}}(E,p) \, \sigma_x  \\
G_{\mathcal{B};21}^{\text{QW}}(E,p) \, \sigma_x   & 0
\end{bmatrix} \, , \label{eq:antidiagG}
\end{align}    
\end{subequations}}
where remember that $\mathcal B \equiv \mathcal B_{\text{QW}}$, and that $G_{\mathcal{B}}^{\text{QW}}(E,p)$ is given by Eq.\ \eqref{eq:final_expression_bis}.

To better visualize what is going on, let us write a bit more explicitly the three integrals we are interested in, on top of each other as before:
{\small
\par
\nobreak
\begin{subequations}
\label{eqs:integrals}
\begin{align}
I^{(1)}(D,\Delta n) &=  2 (-1)^{D+\Delta n} \int_{0}^{\tfrac{\pi}{\e}}  \frac{dp}{2 \pi }  \,e^{ipD\e}   \int_{\tfrac{\pi}{\e}-p}^{\tfrac{\pi}{\e}} \frac{dE}{2 \pi }   \hspace{1mm} e^{-iE\Delta n \e} \nonumber \\
&  \ \ \ \    \times \begin{bmatrix}
G_{\mathcal{B};11}^{\text{QW}}(E,p) \, \mathbb I_{\text{flav.}}  & - G_{\mathcal{B};12}^{\text{QW}}(E,p) \, \sigma_x \\
- G_{\mathcal{B};21}^{\text{QW}}(E,p) \, \sigma_x & G_{\mathcal{B};22}^{\text{QW}}(E,p) \, \mathbb I_{\text{flav.}} 
\end{bmatrix} \\
I^{(2)}(D,\Delta n) &= 2 (-1)^{D+\Delta n} \int_{0}^{\tfrac{\pi}{\e}} \frac{dp}{2 \pi } \, e^{ip D \e} \int_{-\tfrac{\pi}{\e}}^{-\tfrac{\pi}{\e}+p} \frac{dE}{2 \pi }   \hspace{1mm}  
			e^{-iE\Delta n \e} \nonumber \\
&\ \ \ \    \times \begin{bmatrix}
G_{\mathcal{B};11}^{\text{QW}}(E,p) \, \mathbb I_{\text{flav.}} & - G_{\mathcal{B};12}^{\text{QW}}(E,p) \, \sigma_x \\
- G_{\mathcal{B};21}^{\text{QW}}(E,p) \, \sigma_x  & G_{\mathcal{B};22}^{\text{QW}}(E,p) \, \mathbb I_{\text{flav.}} 
\end{bmatrix}   \\
I^{(3)}(D,\Delta n) &= \ \ \ \ \ \ \ \ \ \ \ 2 \ \ \int_{0}^{\tfrac{\pi}{\e}} \frac{dp}{2 \pi } \, e^{ip D \e}  \int_{-\tfrac{\pi}{\e}+p}^{\tfrac{\pi}{\e}-p} \frac{dE}{2 \pi }   \hspace{1mm}  
			e^{-iE\Delta n \e}  \nonumber \\
&\ \ \ \    \times \begin{bmatrix}
G_{\mathcal{B};11}^{\text{QW}}(E,p) \, \mathbb I_{\text{flav.}}   &  + G_{\mathcal{B};12}^{\text{QW}}(E,p) \, \sigma_x \\
 + G_{\mathcal{B};21}^{\text{QW}}(E,p) \, \sigma_x  & G_{\mathcal{B};22}^{\text{QW}}(E,p) \, \mathbb I_{\text{flav.}} 
\end{bmatrix}  \, .
\end{align}    
\end{subequations}}

We see that, when $D+\Delta n$ is even, we get that the sum of the diagonal parts of the three previous integrals is given by
{\small
\par
\nobreak
\begin{equation}
\label{eq:diagpart}
G^{\text{FQW}}_{\text{d.}}\!(t,x ; t',x') \equiv 
2 \int_{0}^{\tfrac{\pi}{\e}}\! \frac{dp}{2 \pi } \, e^{ip D \e} \!  \int_{-\tfrac{\pi}{\e}}^{\tfrac{\pi}{\e}} \!  \frac{dE}{2 \pi }   \hspace{0.5mm}  e^{-iE\Delta n \e} G_{\text{BZ};\text{d.}}^{\text{FQW}}\!(E,p) \, ,
\end{equation}}
where we recall that $G_{\text{BZ};\text{d.}}^{\text{FQW}}\!(E,p)$ is given by Eq.\ \eqref{eq:diagG}. We also realize that when $D+\Delta n$ is odd, we get that the sum of the anti-diagonal parts of the three previous integrals is given by
{\small
\par
\nobreak
\begin{equation}
\label{eq:diagpart}
G^{\text{FQW}}_{\text{n.d.}}\!(t,x ; t',x') \equiv 
2 \int_{0}^{\tfrac{\pi}{\e}}\! \frac{dp}{2 \pi } \, e^{ip D \e} \!  \int_{-\tfrac{\pi}{\e}}^{\tfrac{\pi}{\e}} \!  \frac{dE}{2 \pi }   \hspace{0.5mm}  e^{-iE\Delta n \e} G_{\text{BZ};\text{n.d.}}^{\text{FQW}}\!(E,p) \, ,
\end{equation}}
where we recall that $G_{\text{BZ};\text{n.d.}}^{\text{FQW}}\!(E,p)$ is given by Eq.\ \eqref{eq:antidiagG}.

In App.\ \ref{app:theproof}, we show (i) that the two previous equations, that is, $G^{\text{FQW}}_{\text{d.}}\!(t,x ; t',x')$ when $D+\Delta n$ is even, and $G^{\text{FQW}}_{\text{n.d.}}\!(t,x ; t',x')$ when $D+\Delta n$ is odd, are actually the two only non-vanishing contributions to the total FDQCA GF, together with (ii) the fact that---also in these parity cases for $D+\Delta n$---one has
\begin{subequations}
 \begin{align}
G^{\text{FQW}}_{\text{d.}}\!(t,x ; t',x') &= G^{\text{QW}}_{\text{d.}}\!(t,x ; t',x') \otimes \mathbb{I}_{\text{flav.}} \\ 
G^{\text{FQW}}_{\text{n.d.}}\!(t,x ; t',x') &= G^{\text{QW}}_{\text{d.}}\!(t,x ; t',x') \otimes {\sigma}^1_{\text{flav.}} \, ,
\end{align}   
\end{subequations}
which means that in the end the FDQCA GF reads
\par\nobreak
{\small
\begin{align}
\label{eq:thefinalalternateexpression}
G^{\text{FQW}}(t,x;t',x') = \frac{1}{2} \Big[ (-1)^{D+\Delta n} +1 \Big] G^{\text{QW}}_{\text{d.}}(t,x;t',x') \otimes \mathbb I_{\text{flav.}} \nonumber \\ + \frac{1}{2} \Big[ (-1)^{D+\Delta n +1} +1 \Big] G^{\text{QW}}_{\text{n.d.}}(t,x;t',x') \otimes \mathbb \sigma^1_{\text{flav.}} \, ,
\end{align}}
where we recall that $D\defeq(x-x')/\e$ and $\Delta n \defeq (t-t')/\e$, and where we have used the following obvious notations,
{\small
\par
\nobreak
\begin{subequations}
\begin{align}
G^{\text{QW}}_{\text{d.}}(t,x;t',x') &= \text{diag}\Big(G^{\text{QW}}_{11}(t,x;t',x'), G^{\text{QW}}_{22}(t,x;t',x')\Big) \label{eq:GQWdiag} \\
G^{\text{QW}}_{\text{n.d.}}(t,x;t',x') &= \begin{bmatrix}
0 & G_{12}^{\text{QW}}(t,x;t',x') \\
G_{21}^{\text{QW}}(t,x;t',x') & 0
\end{bmatrix} \, . \label{eq:GQWantidiag} 
\end{align}    
\end{subequations}}
We also give, in App.\ \ref{app:induction}, another proof of Eq.\ \eqref{eq:thefinalalternateexpression}, by induction this time---on $\Delta n$.

Recall that for $\Delta n = 1$, the two previous expressions are respectively given by the diagonal and anti-diagonal parts of Eq.\ \eqref{eq:thefinaleqisthisone}. For an arbitrary $\Delta n$, one could try to compute and, above all, give a simple-enough expression for $G^{\text{QW}}(t,x;t',x')$, either by Fourier integration, or by induction, and then one could insert these expressions into Eq.\ \eqref{eq:thefinalalternateexpression}.

In the present case of the FDQCA, we will not, contrary to what we did do previously for the original QCA, do a comparison between the one-time-step (Fourier-space) GF's of the models $\text{M}=\text{Dirac}$, $\text{M}=\text{FQW}$, \mbox{and $\text{M}=\text{Staggered LGT}$}, because the BZs are too different from each other for these models, whereas in the case of the original QCA, at least the QW and the LGT models had the same integration interval over $p$.

\section{Conclusion and Discussion}
\label{sec:lastsection}

\noindent
Let us discuss our results. It would be of course interesting to compare the $n$-time-step, and not only the one-time-step Green's functions of the Dirac QCA and of continuous-time LGT, with respect to how well they both reach their naive continuum limit and also still manage to approximate it even when far from the continuum limit. And doing this fully in direct space would of course be good---rather than using essentially the Fourier-space expressions as we have done here. Of course, such an extra study should be accompanied by a study of how the FD problem manifests itself at the level of these $n$-time-step Green's functions: even very close to the \emph{naive} continuum limit, the (direct-space, $n$-time-step) Dirac-QCA Green's function will \emph{NOT} match the associated Dirac-equation's Green's function, and this should be visible on the graphical representations of theses functions---again, it would be interesting to see how the fermion-doubling problem manifests itself in such a direct manner, i.e., on these representative curves. The (direct-space, $n$-time-step)  Green's function of the FDQCA, instead, \emph{must} match the associated Dirac-equation's Green's function when we are close to the continuum-limit situation.

Also, this work should be extendable to the $(3+1)$D and $(2+1)$D models of Refs.\ \cite{EDMMplus2023} without much conceptual difficulty, with the help of \mbox{Ref.\ \cite{DAA2025}}.

Finally, we would like to comment on the FD-fixing solution presented in \mbox{Ref.\ \cite{DAA2025}}, called \emph{flavor-staggering-only}, and the necessity of which has been definitively demonstrated in the present article. This solution is minimal, in the sense that it tackles only the doublers that appear in the Dirac QCA, contrary to what would do a direct use of the method used in standard \mbox{LGT \cite{RotheBook}}. More precisley, the $(1+1)$D Dirac QCA only suffers from $1/3$ of the doublers of continuous-time naive lattice fermions---which is the fermionic part of continuous-time QED LGT. Of course, minimality could, naively, be seen as a step towards simplicity. However, there is a non-negligible price we have to pay for this minimality of our solution in terms of both technical and conceptual difficulties: we transform a very standard, square BZ, into a \emph{rhombus} BZ, and this is not without bringing difficulties. In particular, proving, via a proof in direct space (as opposed to a Fourier-space proof), that our FDQCA of \mbox{Ref.\ \cite{DAA2025}} does reach the desired continuum limit, is not easy, because it involves performing a rotation of the coordinate system, and in a discrete spacetime, which requires a lot of care. This has been done in App.\ D of \mbox{Ref.\ \cite{DAA2025}}. All this being said, let us give one very important argument of our flavor-staggering-only solution to fix FD: the fact that we do not stagger chirality implies that our fix preserves the chiral symmetry of the massless original scheme, and that, overall, we actually bypass the Nielsen-Ninomiya no-go theorem; in particular, this has enabled us to suggest, again in \mbox{Ref.\ \cite{DAA2025}}, a model to simulate fermions interacting via the weak force.

\section*{ACKNOWLEDGEMENTS}

\noindent
The authors thank Pablo Arrighi for his occasional insights, and for his support throughout this project.
This work has been partially funded by the French National Research Agency (ANR), (i) within the framework of “Plan France 2030”, under the research
projects EPIQ ANR-22-PETQ-0007, OQULUS ANR-23-PETQ-0013, HQI-Acquisition ANR-22-PNCQ-0001, and HQI-R\&D ANR-22-PNCQ-0002, and (ii) by the project TaQC ANR-22-CE47-0012.

\bibliography{biblio}

\appendix

\section{\emph{Naive} continuum limit of the Dirac-QCA one-time-step Green's function: matches the Dirac-equation Green's function}
\label{app:contlim}

\noindent
If one knows no general mathematical theorem, one can say that there are a priori two nonequivalent ways of checking that our free lattice model reproduces the free Dirac equation in the naive continuum limit $\epsilon \rightarrow 0$, and by this we mean, more specifically: ``reproduces'' the Dirac-equation GF. The first one is to simply evaluate the continuum limit of the EOM operator (see \mbox{Eq.\ \eqref{eq:modifiedEOMoperator}}), and to check that it gives the (Hamiltonian) Dirac-equation EOM operator---after which we can find the GF of the latter. We actually of course already know this is true, since this is the primary very basic reason why we study this model, but we are going to recall these computations---which will also lead us to recall the Dirac-equation GF (in $(1+1)$D spacetime). The second way of checking that our free lattice model reproduces the free Dirac equation in the naive continuum limit is to take the naive continuum limit of the Dirac-QCA one-time-step GF that we have obtained, and to check that it corresponds to the Dirac-equation GF previously recalled. In short: the naive-continuum-limit ``operation'' should commute with the ``taking-the-inverse'' ``operation''---of some EOM operator, be it in discrete spacetime or in continuous sapcetime---if our computational framework behaves well enough. We would have a mathematical procedure that takes as input some discrete-spacetime EOM operator, and that outputs, after taking first the naive continuum limit and then the inverse, or vice versa, the naive-continuum-limit GF. We are going to show explicitly, pedestrianly, that the two ``operations'' commute, since this will make the reader familiar with our framework, and along the way we will make connections with objects known by the QFT-familiar reader. That being said, the commutation of the limit and of taking the inverse is known to hold if the discrete-spacetime EOM operator goes to some (naive-)limit operator in operator norm, and if this limit operator is known to be itself invertible (this result is an application of Proposition 1.2.16 of \mbox{Ref.\ \cite{MelchiorWirth2025notesinFunctionalAnalysisandOperatorTheory}})---provided we disregarded the fact that the naive continuum limit is actually not a proper continuum limit.

\subsection{Dirac-equation Green's function}
\label{app:contlim1}

\noindent
Let us start with the first part announced above. We consider the EOM operator of our Dirac QCA, Eq.\ \eqref{eq:modifiedEOMoperator}. Since by construction 
\begin{equation}
\mathcal T_\epsilon \equiv e^{i \epsilon \mathcal E} \, ,  
\end{equation}
with $\mathcal E \defeq i\partial_t$, and
\begin{equation}
\mathcal S_\epsilon \equiv e^{-i \epsilon \mathcal P_x} \, ,  
\end{equation}
with $\mathcal P_x \defeq - i\partial_x$, then the expansion of Eq.\ \eqref{eq:modifiedEOMoperator} at first order in $\epsilon$ delivers, after some simplifications,
\begin{align}
\mathcal{L}^{\text{QW}}_{ \epsilon} = \begin{bmatrix}
    \mathcal E - \mathcal P_x  & -m \\
    -m & \mathcal E + \mathcal P_x
\end{bmatrix} + O(\epsilon) \, .
\end{align}
Hence,
\begin{equation}
\lim_{\epsilon \rightarrow 0} \Big[\mathcal L^{\text{QW}}_{\epsilon} \Big] =  
\begin{bmatrix}
    \mathcal E - \mathcal P_x & -m \\
    -m & \mathcal E + \mathcal P_x
\end{bmatrix} \, , 
\end{equation}
which can be rewritten
\begin{equation}
\begin{bmatrix}
    \mathcal E - \mathcal P_x & -m \\
    -m & \mathcal E + \mathcal P_x
\end{bmatrix} = \mathcal E - \mathcal H^{\text{Dirac}} \, ,   
\end{equation}
where
\begin{equation}
\label{eq:HDirac}
\mathcal H^{\text{Dirac}} \defeq \alpha^1 \mathcal P_x + m \alpha^0 \, ,
\end{equation}
with
\begin{subequations}
\label{eqs:alphas0}
\begin{align}
\alpha^0 &\defeq \sigma^1 \\
\alpha^1 &\defeq \sigma^3 \, .
\end{align}   
\end{subequations}
We recognize $\mathcal H^{\text{Dirac}}$ to be the one-particle Dirac Hamiltonian in $(1+1)$D dimensions, written in the particular Clifford-algebra representation induced by the alpha matrices of Eqs.\ \eqref{eqs:alphas0}, that is,
\begin{subequations}
\label{eqs:alphas}
\begin{align}
\gamma^0 &\defeq \alpha^0 = \sigma^1 \\
\gamma^1 &\defeq \alpha^0 \alpha^1 = - i \sigma^2 \, .
\end{align}   
\end{subequations}
Hence, the operator
\begin{equation}
\mathcal L^{\text{Dirac}} \defeq   \mathcal E - \mathcal H^{\text{Dirac}} \, ,
\end{equation}
is just a valid (Hamiltonian) Dirac-equation EOM operator, which is what we wanted to check.

Remember that this operator is of course related to the usual Dirac operator $\mathcal D $ via 
\begin{equation}
\mathcal L^{\text{Dirac}} = \gamma^0 \mathcal D \, .
\end{equation}
Remember as well that the Dirac operator can be written
\begin{equation}
\mathcal D = \fsl{\mathcal P} - m \, ,     
\end{equation}
where $\mathcal P \defeq (\mathcal P_0, \mathcal P_1)$, with $\mathcal P_0 \defeq \mathcal E$ and $\mathcal P_1 \defeq -\mathcal P_x$, and where we have used the Feynman slash notation, so that $\fsl{\mathcal P} \equiv \gamma^\mu \mathcal P_\mu = \gamma^0 \mathcal P_0 + \gamma^1 \mathcal P_1$, and hence
\begin{equation}
\mathcal D = \gamma^0 \mathcal P^0 - \gamma^1 \mathcal P^1 - m \, ,     
\end{equation}
since $\mathcal P^0 \defeq \mathcal P_0$ and $\mathcal P^1 \defeq - \mathcal P_1$.
In Fourier space, the Hamiltonian Dirac-equation EOM operator simply reads
\begin{subequations}
\begin{align}
L^{\text{Dirac;FS}}(p^0,p^1) &= \gamma^0 (\fsl{P} - m) \\
&= \sigma^1 (\sigma^1 p^0 - \sigma^1\sigma^3 p^1 - m) \\
&= \begin{bmatrix}
p^0 - p^1 & -m \\
-m & p^0 + p^1
\end{bmatrix} \\
&=  \begin{bmatrix}
E - p_x & -m \\
-m & E + p_x
\end{bmatrix} \, ,
\end{align}
\end{subequations}
where $(p^0 \equiv E, p^1 \equiv p_x \equiv p) \in \mathbb R^2$ is the couple of Fourier variables in the continuum, and $P \defeq (p^0, -p^1)$. 

Still in Fourier space, the GF of the previous EOM operator is just the inverse of the latter, that is,
\begin{subequations}
\begin{align}
G^{\text{Dirac;FS}}_{\text{Ham.}}(E,p_x) &=  \frac{1}{E^2 - p_x^2 - m^2} \begin{bmatrix}
E + p_x & m \\
m & E - p_x
\end{bmatrix} \\ 
&= \frac{1}{E^2 - p_x^2 - m^2} (\fsl{P}+m)\gamma^0 \, .
\end{align}
\end{subequations}
Finally, by definition of the direct-space GF, we then obtain that it is given by
\par\nobreak
{\small
\begin{subequations}
\begin{align}
 G^{\text{Dirac}}_{\text{Ham.}}(t,x;t',x') &\equiv \int_{- \infty}^{+\infty}  \frac{dE}{2 \pi } \int_{- \infty}^{+ \infty}  \frac{dp}{2 \pi } \hspace{1mm}  
			e^{-iE(t-t')+ ip(x-x')} \nonumber \\ 
& \ \ \ \ \times G^{\text{Dirac;FS}}_{\text{Ham.}}(E,p_x) \\
 &= \int_{- \infty}^{+\infty}  \frac{dE}{2 \pi } \int_{- \infty}^{+ \infty}  \frac{dp}{2 \pi } \hspace{1mm}  
			e^{-iE(t-t')+ ip(x-x')} \nonumber \\ 
& \ \ \ \ \times \frac{1}{E^2 - p^2 - m^2} \begin{bmatrix}
E+p & m \\
m & E - p
\end{bmatrix} \, , \label{eq:recall}
\end{align}
\end{subequations}}
which is the result we wished to recall.

\subsection{\emph{Naive} continuum limit of the Dirac-QCA Green's function}

\noindent
The naive continuum limit can only be computed ``off shell'', i.e., on the expression containing an integral over the energies, integral which we do not compute, because once we are on shell the spurious doublers are necessarily taken into account. Take, hence, this ``off-shell'' expression, that is, Eq.\ \eqref{eq:propagator}. One can actually focus on the naive continuum limit of the Fourier-space  GF $G_\mathcal{B}(E,p)$, given in Eq.\ \eqref{eq:final_expression}.

On the one hand, taking the naive continuum limit of the denominator,  $\mathscr{D}(E,p)$, leads, after various computational steps, to the following result,
\begin{equation}
\mathscr{D}(E,p) = E^2 - p^2 - m^2 + O(\epsilon) \, .
\end{equation}
On the other hand, taking the naive continuum limit of the matrix-valued numerator finally results in
\par\nobreak
{\small
\begin{equation}
\begin{bmatrix}
c_\epsilon e^{i \epsilon E} e^{i\epsilon p} - 1 &  i s_\epsilon e^{i \epsilon E} \\
 i s_\epsilon e^{i \epsilon E} & c_\epsilon e^{i \epsilon E} e^{-i\epsilon p} - 1
\end{bmatrix} = 
\begin{bmatrix}
E+p &  m \\  m & E - p
\end{bmatrix} + O(\epsilon) \, .
\end{equation}}

Multiplying the two previous equations, we obtain
\begin{equation}
G_\mathcal{B}(E,p) =  \frac{1}{E^2 - p^2 - m^2} \begin{bmatrix}
E+p & m \\
m & E - p
\end{bmatrix} + O(\epsilon) \, .    
\end{equation}
Inserting the previous equation into Eq.\ \eqref{eq:propagator}, we obtain
\par\nobreak
{\small
    \begin{align}\label{eq:propagator22}
	G(t,x ; t',x') &= \int_{- \pi/\epsilon}^{\pi/\epsilon}  \frac{dE}{2 \pi } \int_{- \pi/\epsilon}^{\pi/\epsilon}  \frac{dp}{2 \pi } \hspace{1mm}  
			e^{-iE(t-t')+ ip(x-x')} \nonumber \\ 
& \ \ \ \ \times \left[\frac{1}{E^2 - p^2 - m^2} \begin{bmatrix}
E+p &  m \\
 m & E - p
\end{bmatrix} + O(\epsilon) \right]  \, ,
    \end{align}}

Hence, taking the limit, we obtain
\par\nobreak
{\small
\begin{align}
\lim_{\epsilon \rightarrow 0} \Big[ G(t,x ; t',x') \Big] &= \int_{- \infty}^{+\infty}  \frac{dE}{2 \pi } \int_{- \infty}^{+ \infty}  \frac{dp}{2 \pi } \hspace{1mm}  
			e^{-iE(t-t')+ ip(x-x')} \nonumber \\ 
& \ \ \ \ \times \frac{1}{E^2 - p^2 - m^2} \begin{bmatrix}
E+p & m \\
m & E - p
\end{bmatrix} \, ,
\end{align}}
which matches Eq.\ \eqref{eq:recall}, and ends this appendix ``successfully''.

\section{Computation of the Dirac-QCA one-time-step Green's function}
\label{app:Dirac-QCA_GF}

\noindent
In this appendix, we wish to compute the direct-space GF $G(t,x;t',x')$ out of the Fourier integral of Eq.\ \eqref{eq:propagator}. This equation can be rewritten
\begin{equation}
\label{eq:first}
 G(t,x;t',x') = \int_{-\pi/\epsilon}^{\pi/\epsilon}  \frac{dp}{(2\pi)^2} e^{i p (x-x')} I(p) \, , 
\end{equation}
where
\begin{equation}
\label{eq:Ipfirst}
I(p) \defeq \int_{-\pi/\epsilon}^{\pi/\epsilon} dE \, J(E;p) \, ,
\end{equation}
with integrand
\begin{equation}
J(E;p) \defeq e^{-iE(t-t')} G_{\mathcal B}(E,p) \, .  
\end{equation}

We are going to compute $I(p)$ by complex integration, via the residue theorem, as it is quite standard when the integrand has poles. First of all, let us perform a change of variable\footnote{By looking at Eq.\ \eqref{eq:final_expression_bis}, the reader may wonder why we have not used the variable $z'\defeq e^{+i \epsilon E }$, which appears naturally in the expression of $G_{\mathcal B}(E,p)$: the reason is mostly ``historical''.
} $E \rightarrow z \defeq e^{-i \epsilon E }$.  As $E$ describes the interval $[-\pi/\epsilon, \pi/\epsilon]$ growingly, $z$ describes the unit circle in the complex plane starting at $z_0=-1$ and ending at the same point, in the \emph{inverse} trigonometric direction, i.e., \emph{clockwise}. Moreover, we have $dz = -i\epsilon \, e^{-i \epsilon E } dE$. 

But before doing the change of variable, we let the reader check that the integral $I(p)$ in Eq.\ \eqref{eq:Ipfirst} can be written as
\begin{equation}
\label{eq:Ip}
I(p) = i\epsilon \int_{-\pi/\epsilon}^{\pi/\epsilon} dE \, e^{-iE\epsilon} e^{-iE\epsilon(\frac{t}{\epsilon} - \frac{t'}{\epsilon})} G^{\text{alt.}}(E,p) \, , 
\end{equation}
where 
\par\nobreak
{\small
\begin{subequations}
\begin{align}
G^{\text{alt.}}(E,p) &\defeq i (M^{\text{QW};\mathcal B}_{\epsilon}(E,p))^{-1}\\
&=  i (\mathscr D^{\text{alt.}}(E,p))^{-1} 
\begin{bmatrix}
c_\epsilon  e^{i\epsilon p} - e^{-i \epsilon E} &  i s_\epsilon \\
 i s_\epsilon & c_\epsilon e^{-i\epsilon p} - e^{-i \epsilon E}
\end{bmatrix} \, ,
\end{align}\end{subequations}}
with 
\begin{equation}
\mathscr D^{\text{alt.}}(E,p) \defeq e^{-2 iE\epsilon} - 2  c_{\epsilon} \cos (p\epsilon) e^{-iE\epsilon} + 1 \, ,  
\end{equation}
is simply ($i$ times) the Fourier-space Green's function of the original EOM operator, $\mathcal M^{\text{QW}}_{\epsilon}$, see \mbox{Eq.\ \eqref{eq:eomop}}, and was simply denoted by $G(E,p)$ in \mbox{Ref.\ \cite{DAA2025}}.

Let us now perform the announced change of variable. We consider the particular case $t-t' = \epsilon$. By performing the change of variable within $I(p)$ in Eq.\ \eqref{eq:Ip}, we finally end up with
\begin{equation}
\label{eq:IpwithIpprime}
I(p) = (i\epsilon) \frac{i}{\epsilon} i I'(p) \, ,
\end{equation}
with
\begin{equation}
I'(p) \defeq \ointclockwise dz f(z;p) \, , 
\end{equation}
where
\begin{equation}
f(z;p) = 
\begin{bmatrix}
 f_{11}(z;p) & f_{12}(z;p) \\
 f_{21}(z;p) & f_{22}(z;p)
\end{bmatrix}  \, ,    
\end{equation}
with
\begin{subequations}
\begin{align}
f_{11}(z;p) &\defeq \frac{c_\epsilon e^{i\epsilon p} z - z^2}{\Delta(z;p)}\\
f_{12}(z;p) &\defeq \frac{i s_\epsilon z }{\Delta(z;p)}  \\
f_{21}(z;p) &\defeq \frac{i s_\epsilon z }{\Delta(z;p)}  = f_{12}(z;p)   \\
f_{22}(z;p)  &\defeq  \frac{c_\epsilon e^{-i\epsilon p} z - z^2}{\Delta(z;p)}  \, ,
\end{align}
\end{subequations}
and
\begin{equation}
\Delta(z;p) \defeq  z^2 - 2c_\epsilon \cos(p\epsilon) z +1    \, .
\end{equation}
We can also write $I'(p)$ as
\begin{equation}
I'(p) = 
\begin{bmatrix}
 I'_{11}(p) & I'_{12}(p) \\
 I'_{21}(p) & I'_{22}(p)
\end{bmatrix}  \, ,    
\end{equation}
where
\begin{equation}
I'_{ij}(p) \defeq  \ointclockwise dz f_{ij}(z;p)  \, .  
\end{equation}
 
Before proceeding to applying the residue theorem, we must find the roots of the denominator $\Delta(z;p)$. The denominator $\Delta(z;p)$ is a polynomial of degree 2 in $z$, so that its roots $z_\pm$ can be found easily:
\begin{equation}
z_\pm \defeq c_\epsilon \cos(p\epsilon) \pm i \sqrt{1 - c_\epsilon^2 \cos^2(p\epsilon)} \, .
\end{equation}

Now, applying the residue theorem for each of the $I'_{ij}$'s taking into account that for each root, the index with respect to the unit circle $\mathcal C$ integrated over in the anti-trigonometric direction is $\text{Ind}_{\mathcal{C}}(z_+) = \text{Ind}_{\mathcal{C}}(z_-) = -1$, we finally obtain, for $I(p)$ of Eq.\ \eqref{eq:IpwithIpprime},
\begin{equation}
 I(p) = (i\epsilon) \frac{i}{\epsilon} i
\begin{bmatrix}
 2i\pi c_{\epsilon} e^{-i\epsilon p} & 2 \pi s_{\epsilon} \\
 2 \pi s_{\epsilon} & 2i\pi c_{\epsilon} e^{i\epsilon p} 
\end{bmatrix}  \, .
\end{equation}

If, eventually, we insert the previous expression into Eq.\ \eqref{eq:first}, we finally obtain, after a simple simplification, Eq.\ \eqref{eq:FFinal}.

\section{Propagator of the Dirac QCA: ``same as'' its Green's function}
\label{app:propagator}

\noindent
Here we do not need complex contour integration because the time integral is not involved in the computation of the propagator, contrary to the case of the GF.

The one-time-step propagator of our Dirac QCA, from $t'$ to $t'+\epsilon$, is defined, as usual in quantum mechanics, by
\begin{equation}
 K(t'+\epsilon,x;t',x') \defeq \langle x | \hat{U}_\epsilon^{\text{QW}} | x' \rangle \, ,
\end{equation}
which, by inserting the momentum closure relation, given in Eq.\ \eqref{eq:fermetureFourier}, yields
\par\nobreak
{\small
\begin{subequations}
\begin{align}
 K(t'+\epsilon,x;t',x') &\defeq \langle x | \hat{U}_\epsilon^{\text{QW}} \int_{-\pi/\epsilon}^{\pi/\epsilon} \frac{dp}{2\pi} \ket p \! \bra p | x' \rangle \\
 &= \int_{-\pi/\epsilon}^{\pi/\epsilon} \frac{dp}{2\pi} \, \langle x | \begin{bmatrix}
 c_{\epsilon} \hat S & - i s_\epsilon \\
  - i s_\epsilon & c_{\epsilon} \hat S^\dag
 \end{bmatrix} \ket p e^{-ipx'} \\
&= \int_{-\pi/\epsilon}^{\pi/\epsilon} \frac{dp}{2\pi}  \begin{bmatrix}
 c_{\epsilon} \langle x | e^{-i\epsilon \hat p} \ket p  & - i s_\epsilon \langle x | p \rangle \\
  - i s_\epsilon \langle x | p \rangle & c_{\epsilon}\langle x | e^{i\epsilon \hat p} \ket p    \\
 \end{bmatrix} e^{-ipx'} \\
&= \int_{-\pi/\epsilon}^{\pi/\epsilon} \frac{dp}{2\pi} 
\begin{bmatrix}
c_{\epsilon} e^{ipx} e^{-i\epsilon p} & - i s_\epsilon e^{ipx} \\
- i s_\epsilon e^{ipx}   &   c_{\epsilon} e^{ipx} e^{i\epsilon p} 
\end{bmatrix}  e^{-ipx'}  \\
&= \int_{-\pi/\epsilon}^{\pi/\epsilon} \frac{dp}{2\pi} \, e^{ip(x-x')} 
\begin{bmatrix}
c_{\epsilon} e^{-i\epsilon p} & - i s_\epsilon \\
- i s_\epsilon   &   c_{\epsilon} e^{i\epsilon p}
\end{bmatrix} \, .
\end{align}    
\end{subequations}}
Notice that such a computation by inserting a momentum closure relation turns out to be possible only because we are here dealing with a \emph{free}, i.e., translationally invariant scheme, which is diagonal in the momentum variable. If we were dealing with an interacting model, we would have had to use the standard generic method of diagonalizing the (one-time-step) evolution operator and inserting a closure relation expressed on the eigenbasis of that evolution operator.

\section{Final expression for the direct-space one-time-step Green's function of the Dirac QCA}
\label{app:Finalexpr}

\noindent
Let us perform the integral of Eq.\ \eqref{eq:FFinal}. There are actually $4$ number-valued integrals to perform, since the original one is $4\times 4$-matrix-valued. Indeed, Eq.\ \eqref{eq:FFinal} can be written
\par\nobreak
{\small
\begin{equation}
\label{eq:Gfirst}
G(t'+\epsilon,x;t',x') = \begin{bmatrix}
c_{\epsilon} E_{11}(\epsilon,x-x') & - i s_{\epsilon} E_{12}(\epsilon,x-x') \\
- i s_{\epsilon} E_{21}(\epsilon,x-x') & c_{\epsilon} E_{22}(\epsilon,x-x')
\end{bmatrix} \, ,  
\end{equation}}
with 
\begin{subequations}
\begin{align}
E_{11}(\epsilon,x-x') &= \Sigma(\epsilon,D-1) \\
E_{12}(\epsilon,x-x') = E_{21}(\epsilon,x-x') &=   \Sigma(\epsilon,D) \\
E_{22}(\epsilon,x-x') &= \Sigma(\epsilon,D+1) \, ,
\end{align} 
\end{subequations}
where 
\begin{equation}
    D \defeq \frac{x-x'}{\epsilon} \in \mathbb Z \, ,
\end{equation}
and
\begin{equation}
\Sigma(\epsilon,D) \defeq \int_{-\frac{\pi}{\epsilon}}^{\frac{\pi}{\epsilon}} \frac{dp}{2 \pi} \, e^{ipD\epsilon} \, ,
\end{equation}
with $D \in \mathbb Z$.

Now, when $D\neq0$, the previous integral can easily be shown to vanish, and when $D = 0$, its value is $1/\epsilon$, so that in the end, we recover the following well-known result:
\begin{equation}
\Sigma(\epsilon,D) = \frac{1}{\epsilon} \delta_{x,x'} \, .   
\end{equation}

Using the previous result, we can finally give, \mbox{for $G(t'+\epsilon,x;t',x')$}, in Eq.\ \eqref{eq:Gfirst} above, the expression of Eq.\ \eqref{eq:thefinaleqisthisone}.

\section{A short summary about Green's functions}
\label{app:GFtheory}

\subsection{Introduction}

\noindent
In the context of mathematically classical fields---i.e., these fields are functions of spacetime that are number-valued, as opposed to operator-valued fields as in QFT for example---, there are essentially two principal types of situations in which GFs are useful---i.e., two paradigmatic situations. The first one is the case of \emph{inhomogeneous} linear partial differential equations (PDEs)---that is, involving a field called ``source field'', which is ``external'', that is, a given of the problem, i.e., it is different from the unknown fields for which we seek a solution; in this first paradigmatic situation, we consider the  boundary conditions to be homogeneous---that is, typically, the value of an unknown field (Dirichlet boundary condition), xor/and of its derivatives (Neumann boundary condition), is/are constrained to vanish on the boundaries. The second type of situation in which GFs are useful is for \emph{homogeneous} linear PDEs (the opposite of ``inhomogeneous'') with, this time, inhomogeneous boundary conditions---that is, typically, on the boundaries the value of an unknown field is arbitrary, but of course most often fixed, otherwise the problem of finding a solution is ill-defined/incomplete mathematically if we are looking for a unique solution\footnote{That is to say, we must define a \emph{Cauchy problem} for the Cauchy-Lipschitz theorem to ensure unicity.}. Of course one can combine these two paradigmatic situations into an inhomogeneous linear PDE with inhomogeneous boundary conditions.

The two principal types of situation described above can be seen in a unified manner in the theory of GFs, that is, more or less formally, as ``input data'' to the problem whose central ingredient is the homogeneous part of the PDE itself: in the first type, this input data is the external, source field, while in the second type, this input are the inhomogeneous boundary conditions. The simplest formal way to achieve this ``unification'', is known as Duhamel's principle: the solution of a homogeneous PDE with inhomogeneous boundary conditions can be obtained as the solution of the inhomogeneous PDE that has been obtain by taking the previous homogeneous part, and by putting as a source term, in an appropriate manner, the boundary conditions multiplied by some appropriate Dirac delta functions and appropriate constant prefactors.

\subsection{Continuous-spacetime Schrödinger-like equations}

\subsubsection{Cauchy problem}

\noindent
Let us get into technicalities and into the actual solutions of these problems in the case of Schrödinger-like equations. We want to solve the following equation,
\begin{equation}
\label{eq:theeqtosolve}
 \mathcal Lu |_{t,x} = 0 \, ,   
\end{equation}
where
\begin{equation}
\mathcal L \defeq i \partial_t - \mathcal H \, ,    
\end{equation}
is a linear operator, with $\mathcal H$ an a priori arbitrary Hamiltonian, that is, self-adjoint operator, that we will consider time independent for simplicity, and that may involve derivatives with respect to $x$, namely, $\partial_x$, to an arbitrary power. Equation \eqref{eq:theeqtosolve} is qualified as homogeneous because it involves no other function than the unknown to be determined, and in particular this formally translates into the fact that the RHS of that equation is zero. We choose the following initial condition for our PDE,
\begin{equation}
    u(t_0,x)=u_0(x) \, ,
\end{equation}
qualified as inhomogeneous when $u_0(x) \neq 0$. We now have a well-posed Cauchy problem.

\subsubsection{Green's function resolution method and definition equation}

\noindent
A standard way of solving a linear PDE like \mbox{Eq.\ \eqref{eq:theeqtosolve}} is via the GF method, which we are going to recall step by step in a simple manner, adapted to our present Schrödinger-like problem.

The standard GF definition equation is the following,
\begin{equation}
\label{eq:GFdeff}
 \mathcal L G(\cdot,\cdot;t',x')|_{t,x} = c \, \delta(t-t') \delta(x-x') \, ,  
\end{equation}
where $G(t,x;t',x')$ is the GF to be determined, and where $c \in \mathbb C$ is some constant to be determined, and can be considered as a matter of convenience, since it could be included in a redefinition of $G$.

\subsubsection{Announcing Duhamel's ``principle'', and generic formal solution for the Green's function}

\noindent
Now, in our opinion, the simplest and most enlightening way to understand the use of the Green's function in the case of a homogeneous equation, is via so-called Duhamel's principle, evoked in the previous \mbox{subsection (Introduction)}. Let us then consider the case of an inhomogeneous linear PDE based on the previous homogeneous one:
\begin{equation}
\label{eq:theeqtosolve2}
 \mathcal L u |_{t,x} = f(t,x) \, ,   
\end{equation}
where $f(t,x)$ is called the source term, and is a known, imposed function. Let us now recall that standard observation that makes a GF defined by Eq.\ \eqref{eq:GFdeff} a solution to the previous equation. By multiplying Eq.\ \eqref{eq:GFdeff} by $f(t',x')$ on both sides, and then by integrating over both $x'$ and $t'$ (in a naive manner regarding $t'$, which we will discuss afterwards), we finally arrive to 
\begin{equation}
\mathcal L \left[ \frac{1}{c} \int_{-\infty}^{+\infty} dt' \int_{-\infty}^{+\infty} dx' G(\cdot,\cdot;t',x') f(t',x')  \right]_{t,x} = f(t,x) \, .    
\end{equation}
By comparing the previous equation to the equation we want to solve, Eq.\ \eqref{eq:theeqtosolve2}, we realize that the following formula provides a solution to the latter---in the case of homogeneous, that is, vanishing boundary conditions (we will not recall all these aspects here)---:
\begin{equation}
\label{eq:solution}
u_G(t,x) \defeq \frac{1}{c} \int_{-\infty}^{+\infty} dt' \int_{-\infty}^{+\infty} dx' G(t,x;t',x') f(t',x') \, .
\end{equation}

\subsubsection{Going back to the Schrödinger-like Cauchy problem---which is a homogeneous PDE with inhomogeneous boundary conditions---and solving it}

\noindent
Let us now go back to the initial homogeneous equation, and see how we can make the previous analysis of the inhomogeneous equation fit that of the homogeneous one. Suppose we choose, in the inhomogeneous problem,
\begin{equation}
\label{eq:choice}
f(t',x') = c \, \delta(t_0-t') u_0(x') \, .    
\end{equation}
Let us work out the resulting $u_G$, which we call $u_{G,u_0}$. After some manipulations, we get
\par\nobreak
{\small
\begin{align}
u_{G,u_0}(t,x) = \int_{-\infty}^{+\infty} dx' \left( \int_{-\infty}^{+\infty} dt' \delta(t_0-t') G(t,x;t',x') u_0(t',x') \right) \, ,
\end{align}}
which, after performing the integral over $t'$, results in
\begin{equation}
u_{G,u_0}(t,x) = \int_{-\infty}^{+\infty} dx' G(t,x;t_0,x') u(t_0,x') \, .
\end{equation}
But this is just the well-known ``propagation'' integral equation satisfied by a quantum-mechanical propagator; in other words, a solution for $G$ seems to be just $G=K$, where $K$ is the propagator of our initial Schrödinger-like equation.

\subsubsection{Final crucial comment for complete exactness of what preceeds}

\noindent
Although this is essentially ``it'', there are some subtleties, which we are going to discuss now.  Actually, it turns out that $K$ is not a valid Green's function for \mbox{any $t_0$ and $t$,} i.e., it does not satisfy the Green's function definition equation for any $t_0$ \mbox{and $t$.} In \mbox{Ref.\ \cite{Littlejohn}}, it is shown that one possible Green's function that is valid for any $t_0$ and $t$, is rather \mbox{$G(t,x;t_0,x') = \Theta(t-t_0) K(t,x;t_0,x')$,} where $\Theta(X)$ is the step function, $\Theta(X)=1$ \mbox{if $X>0$,} and $\Theta(X)=0$ otherwise. Hence, the ``only'' discussion remaining in this appendix could be how to ``naturally'' come up with the fact that one valid Green's function for any $t_0$ and $t$ is indeed \mbox{$G(t,x;t_0,x') = \Theta(t-t_0) K(t,x;t_0,x')$,} i.e., how to explain,  just through the previous computations only, that $K(t,x;t_0,x')$ can actually not provide a solution for $G(t,x;t_0,x')$, but that $\Theta(t-t_0) K(t,x;t_0,x')$ can.

\subsubsection{Comments on the constant $c$}

\noindent
Finally, let us discuss the question of the \mbox{constant $c$.} In the previous computation, it seems that in the end the choice of $c$ is completely irrelevant to the final \mbox{results---leaving} alone the fact that anyways, \mbox{changing $c$ in} the GF definition equation (without changing it in Eq.\ \eqref{eq:choice}) would only modify the solution obtained for the GF by a global phase, which in quantum mechanics usually does not matter. Let us have a look at this more in detail. In Eq.\ \eqref{eq:solution}, the choice of $c$ does matter. But then, by choosing $f(t',x')$ appropriately, i.e., essentially, as containing as well a factor of $c$, one can make the choice of $c$ become irrelevant in the final solution. 

That being said, one must pay attention to the following thing. In our example, one would have to justify---in case the following is true, which is actually \mbox{unlikely---why} choosing $f(t',x')$ as in Eq.\ \eqref{eq:choice} is indeed the ``correct'' thing to do when transforming the homogeneous equation with inhomogeneous spacetime-boundary conditions into an inhomogeneous equation with homogeneous spacetime-boundary conditions. Indeed, if one chooses instead $\delta(t_0-t') u_0(x')$, one would in the end find $G = (1/c) K$/ Also (let us extend our analysis to discrete spacetime) the propagation equation of Eq.\ \eqref{eq:propagationequation} would have to be modified. In any case, again, the question one must answer is: which one of Eq.\ \eqref{eq:choice} or of $\delta(t_0-t') u_0(x')$ is the ``correct'' or at least ``natural'' choice for $f(t',x')$. In standard QM, it is actually often the second choice that is made, with $c=i$ \cite{Littlejohn}.

\section{[QFT-QM]-hybrid notations for continuous- and discrete-spacetime QFTs and their respective single-particle sectors formalized as QM theories}
\label{app:hybridconvention}

\subsection{Continuous-spacetime theories}

\noindent
The aim of this subappendix is simple, and the following: set, or recall, the abstract-QM formalism (i.e., with position and momentum kets and bras) that is adapted to describe the single-particle sectors of usual QFTs, the latter being written with the most widespread conventions.

Le us consider a ``classical'' field theory in $1+1$ spacetime dimensions, for some matter field $\psi : \mathbb R^2 \rightarrow \mathbb{C}^N$, possibly multicomponent, i.e., $N \in \mathbb N^\ast$ can be strictly bigger than $1$. The space and time coordinates are respectively denoted by $x\in \mathbb R$ and $t\in \mathbb R$. 

In mathematically classical field theory and hence QFT in $1+1$ spacetime dimensions, the most widespread convention for the Fourier transform in space is
\begin{align}
\tilde{\psi}_{\text{mom.}}(t,p) &\defeq \int_{-\infty}^{+\infty} dx \, \psi(t,x) \, e^{-ipx} \\
\psi(t,x) &= \int_{-\infty}^{+\infty} \frac{dp}{2\pi} \, \tilde{\psi}_{\text{mom.}}(t,p) \, e^{ipx} \, ,
\end{align}
where of course the second equality is implied by the first, which is the definition of the Fourier transform, where all of the convention is chosen. Following this convention, the most widespread convention for the \emph{formal} Fourier transform in both space \emph{and time} is
\begin{subequations}
\label{eqs:spacetimeFTs}
\begin{align}
\tilde{\psi}(E,p) &\defeq \int_{-\infty}^{+\infty} dt  \int_{-\infty}^{+\infty} dx \, \psi(t,x) \, e^{iEt -ipx} \\
\psi(t,x) &= \int_{-\infty}^{+\infty} \frac{dE}{2\pi} \int_{-\infty}^{+\infty} \frac{dp}{2\pi} \, \tilde{\psi}(E,p) \, e^{-iEt + ipx} \, ,
\end{align}
\end{subequations}
with the same comment as for the spatial Fourier transform above.

Let us now introduce the abstract-QM formalism that is adapted to the previous ``classical''-field-theory conventions. First of all, as always, we introduce $\ket{\psi(t)}$, the abstract-QM state of the formalism, which is related to the wavefunction by a standard relation that we will recall below. The position and momentum bases $\{ \ket{x} : x \in \mathbb R \}$ and $\{ \ket{p} : p \in \mathbb R \}$, respectively, of the position abstract Hilbert space $\mathcal H_{\text{pos.}}$ of our abstract-QM formalism, are such that\footnote{In the context of the most standard connections of QM to QFT, one sometimes have  (for in example in \mbox{Ref.\ \cite{book_Schwartz}}, Eq.\ (2.73)),  for the momentum closure relation, $\int_{-\infty}^{\infty} \frac{dp}{2\pi } \frac{1}{2 \omega(p)}  \ket p \! \bra p = \mathbb I$, where $\omega(p)$ is the dispersion relation of the free model. Here, we could have done the same, by including the energy (or frequency) branches $\omega^{\text{QW}}_b(p)$ of the QW model, and more generally, $\omega^{\text{M}}_b(p)$ for any free model M, with $b$ the branch index.}
\begin{subequations}
\label{eqs:closurerelationsXP}
\begin{align}
\int_{-\infty}^{\infty} dx \ket x \! \bra x &= \mathbb I \\
\int_{-\infty}^{\infty} \frac{dp}{2\pi} \ket p \! \bra p &= \mathbb I \, ,
\label{eq:fermetureFourier}
\end{align}
\end{subequations}
and
\begin{equation}
\langle x | p \rangle = e^{ipx} \, .
\label{eq:scalarproduct}
\end{equation}
As always, the wavefunction is by construction related to the abstract-QM state via
\begin{equation}
\psi(t,x) \equiv \langle x | \psi (t) \rangle \, ,
\end{equation}
and the reader can check that Eqs.\ \eqref{eq:fermetureFourier} and \eqref{eq:scalarproduct} imply that
\begin{equation}
\langle p | \psi (t) \rangle = \tilde{\psi}_{\text{mom.}}(t,p) \, .
\end{equation}

Similarly, it can also be useful to introduce formal time and energy abstract bases $\{ \ket{t} : t \in \mathbb R \}$ and $\{ \ket{E} : E \in \mathbb R \}$, respectively, of some enlarged formal spacetime-position Hilbert space $\mathcal{H}_{\text{full}}$, whose elements are states $\ket{\psi}$ encoding the full knowledge of $\psi(t)$, via
\begin{equation}
\psi(t) \equiv \langle t | \psi \rangle \, .    
\end{equation}
The time and energy bases are such that
\begin{subequations}
\label{eqs:closurerelationsTE}
\begin{align}
\int_{-\infty}^{\infty} dt \ket t \! \bra t &= \mathbb I \\
\int_{-\infty}^{\infty} \frac{dE}{2\pi} \ket E \! \bra E &= \mathbb I \, ,
\label{eq:fermetureFourierE}
\end{align}
\end{subequations}

\subsection{Discrete-spacetime theories}

\noindent
When one wants to spacetime discretize a continuum quantum-mechanical theory, or theory of ``classical'' fields---in the latter case, often with the perspective of spacetime discretizing a quantum field theory---, the main issue is that of the scaling of the discrete-spacetime-theory objects with the spacetime-lattice spacing.

In natural units ``$\hbar=c=1$'', we can in particular consider, not only that ``$\hbar = 1 \, \text{unit of an action}$'', but also, only as a mnemotechnic means, but which is more physically meaningful than speaking about units, that actions are dimensionless---rather, that we choose actions to be the no-dimension \emph{reference}---, i.e., in particular, $[\hbar] = 1$, so that in the end we simply have $\hbar = 1$ (without units).

Before proceeding, let us make a remark about the continuum theory. Although each $\psi_i$, $i=1,...,N$, takes, at each spacetime point, numerical values in $\mathbb C$, the modulus of $\psi_i$ is however not dimensionless, i.e., its numerical value is in certain physical units, in such a way that the following normalization integral is, this time, dimensionless, since it must correspond to a probability: $\forall t$,
\begin{equation}
    \int_{-\infty}^{+\infty} dx \, |\psi(t,x)|^2 = 1 \, .
\end{equation}
The previous equation implies that the dimension of the modulus of $\psi(t,x)$ must be that of the inverse of the square root of a length, the latter being often denoted \mbox{as $L^{\frac{1}{2}}$.}

Let us now start our spacetime discretization from the remark of the previous paragraph. The straightfoward Riemann-sum discretization of the previous equation is simply
\begin{equation}
    \sum_{x \in \mathbb Z \epsilon} \epsilon \, |\psi(t,x)|^2 = 1 \, ,
\end{equation}
where this time both $t$ and $x$ belong to $\mathbb Z \epsilon$.
The previous equation shows that, if we want dimensionless wavefunctions in order to be able to work only with dimensionless numbers  all across our discrete-spacetime theory and for all the objects of our theory, we need to define some
\begin{equation}
\psi_{\text{latt.}}(t,x) \defeq \sqrt{\epsilon} \psi(t,x) \, , 
\end{equation}
which is the dimensionless wavefunction announced. Now, the point of the present particular [QFT-QM]-hybrid notation for \emph{discrete}-spacetime field theories that we want to put forward is, precisely, \emph{not} to use dimensionless wavefunctions, in order to be able to take naive continuum limits more easily\footnote{Notice, however, that easy continuum limits can only be taken in free theories, but this is already an interesting-enough check for us to adapt the framework to taking such limits. In interacting theories, continuum limits are more involved, not to speak of \emph{quantum} interacting field theories.}.

Hence, we will use the following spacetime discretization of Eqs.\ \eqref{eqs:spacetimeFTs},
\begin{subequations}
\label{eqs:spacetimeFTsDISCRETE}
\begin{align}
\tilde{\psi}(E,p) &\defeq \sum_{t \in \mathbb Z \epsilon} \epsilon \sum_{x \in \mathbb Z \epsilon} \epsilon \, \psi(t,x) \, e^{iEt -ipx} \\
\psi(t,x) &= \int_{-\frac{\pi}{\epsilon}}^{\frac{\pi}{\epsilon}} \frac{dE}{2\pi} \int_{-\frac{\pi}{\epsilon}}^{\frac{\pi}{\epsilon}} \frac{dp}{2\pi} \, \tilde{\psi}(E,p) \, e^{-iEt + ipx} \, .
\end{align}
\end{subequations}
These formulae match standard conventions and formulae for Fourier transforms of sequences defined on infinite lattices \cite{book_Chu2008}.
When it comes to the spacetime discretization of Eqs.\ \eqref{eqs:closurerelationsXP} and \eqref{eqs:closurerelationsTE}, it yields, respectively,
\begin{subequations}
\begin{align}
\sum_{x \in \mathbb Z \epsilon} \epsilon \ket x \! \bra x &= \mathbb I \\
\int_{-\frac{\pi}{\epsilon}}^{\frac{\pi}{\epsilon}} \frac{dp}{2\pi} \ket p \! \bra p &= \mathbb I \, ,
\label{eq:fermetureFourier1bis}
\end{align}
\end{subequations}
and
\begin{subequations}
\begin{align}
\sum_{t \in \mathbb Z \epsilon} \epsilon \ket t \! \bra t &= \mathbb I \\
\int_{-\frac{\pi}{\epsilon}}^{\frac{\pi}{\epsilon}} \frac{dE}{2\pi} \ket E \! \bra E &= \mathbb I \, ,
\label{eq:fermetureFourier2}
\end{align}
\end{subequations}

Finally, let us mention one last important thing. The Dirac delta distribution used for continuous variables, say, $\delta(x-x_0)$, must be replaced, in discrete spacetime, by $\frac{1}{\epsilon} \delta_{x,x_0}$, where $\delta_{x,x_0}$ is the Kronecker delta symbol.

\section{Taylor expansion of the direct-space one-time-step Dirac-QCA Green's function}
 \label{app:TaylorExpansion}

\noindent
Since we are going to perform a Taylor expansion in $\epsilon$, let us use the continuum notation for $(1/\epsilon)\delta_{x',x}$, which is $\delta(x'-x)$. This yields, for Eq.\ \eqref{eq:thefinaleqisthisone}, the following expression,
\begin{subequations}
\begin{align}
G(t'+\epsilon,x;t',x') &=  \left[\begin{array}{cc}
 c_\epsilon \delta(x'- (x-\epsilon)) & - i s_\epsilon \delta(x'- x)  \\
 - i s_\epsilon \delta(x'- x)  &  c_\epsilon \delta(x'- (x+\epsilon))
\end{array}  \right] \\
&= \left[\begin{array}{cc}
 c_\epsilon \delta((x-\epsilon) - x') & - i s_\epsilon \delta(x- x')  \\
 - i s_\epsilon \delta(x- x')  &  c_\epsilon \delta((x+\epsilon) - x'))
\end{array}  \right] \\
&= \left[\begin{array}{cc}
 c_\epsilon \delta((x-x') - \epsilon) & - i s_\epsilon \delta(x- x')  \\
 - i s_\epsilon \delta(x- x')  &  c_\epsilon \delta((x-x') + \epsilon)
\end{array}  \right] \, .
\end{align} 
\end{subequations}
We are now going to perform the announced Taylor expansion in $\epsilon$, around $0$ for the cosines and sines, and around $x-x'$ for the delta functions. This gives
\begin{widetext}
\begin{subequations}
 \begin{align}
G(t'+\epsilon,x;t',x')  &=\left[\begin{array}{cc}
\delta(x-x') - \epsilon  \partial_X\delta|_{X=x-x'} & -i \epsilon m  \, \delta(x-x') \\
-i\epsilon m \, \delta(x-x') &   \delta(x-x') + \epsilon  \partial_X\delta|_{X=x-x'}
\end{array}  \right]  + O(\epsilon^2) \\
&= \left[\begin{array}{cc}
\delta(x-x') - \epsilon  \delta'(x-x') & -i \epsilon m  \, \delta(x-x') \\
-i\epsilon m \, \delta(x-x') &   \delta(x-x') + \epsilon  \delta'(x-x')
\end{array}  \right]  + O(\epsilon^2)\\
&= \delta(x-x') + \epsilon \left[\begin{array}{cc}
- \delta'(x-x') & -i  m  \, \delta(x-x') \\
-i m \, \delta(x-x') &  \delta'(x-x')
\end{array}  \right] + O(\epsilon^2)
\end{align}    
\end{subequations}

Let us now insert the previous expression in the ``propagation equation'',  Eq.\ \eqref{eq:propagationequation}. This reads
\begin{align}
\psi(t'+\epsilon,x) = \int dx' \left( \delta(x-x') + \epsilon \left[\begin{array}{cc}
- \delta'(x-x') & -i \epsilon m  \, \delta(x-x') \\
-i\epsilon m \, \delta(x-x') &  \delta'(x-x')
\end{array}  \right] \right) \psi(t',x') + O(\epsilon^2)\, ,   
\end{align}    
which, after a Taylor expansion of the LHS, and some manipulations on the RHS, gives
\begin{align}
\psi(t',x) + \epsilon \partial_{0}\psi|_{t',x}  &= \psi(t',x) + \epsilon  \left(\begin{array}{cc}
- \int dx' \delta'(x-x') \psi^+(t',x')  -i m \int dx'  \, \delta(x-x') \psi^-(t',x') \\
-i  m  \int dx'  \delta(x-x') \psi^+(t',x')  + \int dx'   \delta'(x-x')\psi^-(t',x')
\end{array}  \right) + O(\epsilon^2) \, .
\end{align}    
The zeroth orders cancel each other, we can then divide by $\epsilon$, let $\epsilon$ go to zero, and perform the trivial ``off-diagonal'' integrals, which gives
\begin{align}
\label{eq:stillthedeltaprimes}
\partial_{0}\psi|_{t',x}  &= \left( \begin{array}{cc}
- \int dx' \delta'(x-x') \psi^+(t',x')  -i m \, \psi^-(t',x) \\
-i  m \, \psi^+(t',x)  + \int dx'   \delta'(x-x')\psi^-(t',x')
\end{array}  \right) \, .
\end{align} 
\end{widetext}
We are left with the integrals involving the derivative of the delta function. Regarding those, we can write the following general result,
\begin{subequations}
\begin{align}
\int dx' \delta'(x-x') \psi^i(t',x') &= \int dx' \delta'(\underbrace{x'-x}_{x''}) \psi^i(t',x') \\
&= \int dx'' \delta'(x'') \psi^i(t',x''+x) \\
&= \partial_x \psi^i|_{t',x} \, ,   
\end{align}    
\end{subequations}
which, used in Eq.\ \eqref{eq:stillthedeltaprimes}, delivers, using the \mbox{notation $\partial_0 = \partial_t$},
\begin{align}
\partial_t \psi =  \left( \begin{array}{cc}
- \partial_x \psi^+  -i m \, \psi^- \\
\partial_x \psi^- -i  m \, \psi^+ 
\end{array}  \right)   \, ,
\end{align}
which, multiplied by $i$, gives
\begin{equation}
i \partial_t \psi = \mathcal{H}^{\text{Dirac}} \psi \, , 
\end{equation}
where $\mathcal{H}^{\text{Dirac}}$ is given in Eq.\ \eqref{eq:HDirac}.

\section{Integrands of the one-time-step Green's functions}
\label{app:theintegrands}

\noindent
For the Dirac, continuum model, one way of writing the relevant expressions is
\begin{subequations}
\begin{align}
\mathrm{Re}\{ I^{\text{Dirac}}_{11}(p;\epsilon,\Delta) \} &= \frac{e^{ip\Delta}}{2\pi}  a^{\text{Dirac}}(p;\epsilon) \\
&= \frac{e^{ip\Delta}}{2\pi}   \cos(\epsilon\sqrt{p^2+m^2}) \\
- \mathrm{Im}\{ I^{\text{Dirac}}_{11}(p;\epsilon,\Delta) \} &= \frac{e^{ip\Delta}}{2\pi}  b^{\text{Dirac}}(p;\epsilon) \\
&= \frac{e^{ip\Delta}}{2\pi} \epsilon p \, \mathrm{sinc}(\epsilon \sqrt{p^2+m^2}) \\
- \mathrm{Im}\{ I^{\text{Dirac}}_{i,j\neq i}(p;\epsilon,\Delta) \} &= \frac{e^{ip\Delta}}{2\pi}  c^{\text{Dirac}}(p;\epsilon) \\
&= \frac{e^{ip\Delta}}{2\pi} \epsilon m \, \mathrm{sinc}(\epsilon \sqrt{p^2+m^2}) \, .
\end{align}    
\end{subequations}
For the LGT model, we have
\begin{subequations}
\begin{align}
\mathrm{Re}\{ I^{\text{LGT}}_{11}(p;\epsilon,\Delta) \} &= \frac{e^{ip\Delta}}{2\pi}  a^{\text{LGT}}(p;\epsilon) \\
&= \frac{e^{ip\Delta}}{2\pi} \cos\left(\epsilon \sqrt{\sin^2(p)+m^2}\right) \\
- \mathrm{Im}\{ I^{\text{LGT}}_{11}(p;\epsilon,\Delta) \} &= \frac{e^{ip\Delta}}{2\pi}  b^{\text{LGT}}(p;\epsilon) \\
&= \frac{e^{ip\Delta}}{2\pi} \sin(\epsilon p) \, \mathrm{sinc}\left(\epsilon\sqrt{\sin^2(p)+m^2}\right)  \\
- \mathrm{Im}\{ I^{\text{LGT}}_{i,j\neq i}(p;\epsilon,\Delta) \} &= \frac{e^{ip\Delta}}{2\pi}  c^{\text{LGT}}(p;\epsilon) \\
&= \epsilon m \, \mathrm{sinc}\left(\epsilon\sqrt{\sin^2(p)+m^2} \right) \, .
\end{align}    
\end{subequations}
For the QW model, we have
\begin{subequations}
\begin{align}
\mathrm{Re}\{ I^{\text{QW}}_{11}(p;\epsilon,\Delta) \} &= \frac{e^{ip\Delta}}{2\pi}  a^{\text{QW}}(p;\epsilon) \\
&= \frac{e^{ip\Delta}}{2\pi} \cos(\epsilon m)\cos(\epsilon p)  \\
- \mathrm{Im}\{ I^{\text{QW}}_{11}(p;\epsilon,\Delta) \} &= \frac{e^{ip\Delta}}{2\pi}  b^{\text{QW}}( p;\epsilon) \\
&= \frac{e^{ip\Delta}}{2\pi} \cos(\epsilon m)\sin(\epsilon m)  \\
- \mathrm{Im}\{ I^{\text{QW}}_{i,j\neq i}(p;\epsilon,\Delta) \} &= \frac{e^{ip\Delta}}{2\pi}  c^{\text{QW}}(p;\epsilon) \\
&= \frac{e^{ip\Delta}}{2\pi} \sin(\epsilon m) \, .
\end{align}    
\end{subequations}

\section{Plotting the $B^{\mathrm{M}}(P;M)$'s and the $C^{\mathrm{M}}(P;M)$'s}
\label{app:otherfigures}

\begin{figure*}
    \centering
    \includegraphics[width=0.8\linewidth]{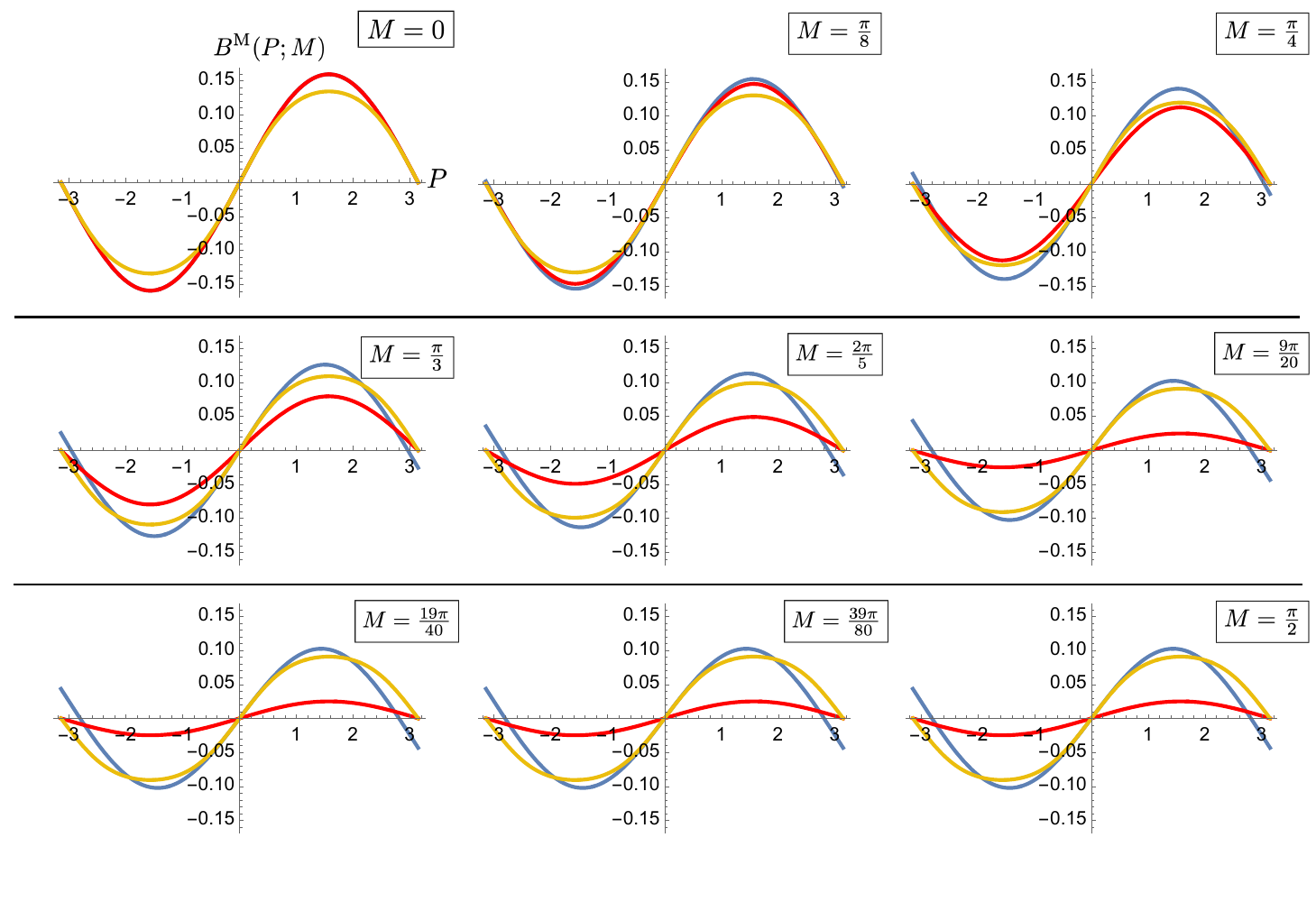}
    \caption{Same as Fig.\ \ref{fig:As} but for the $B^{\text{M}}(P;M)/(2\pi)$'s.}
    \label{fig:Bs}
\end{figure*}

\begin{figure*}
    \centering
    \includegraphics[width=0.8\linewidth]{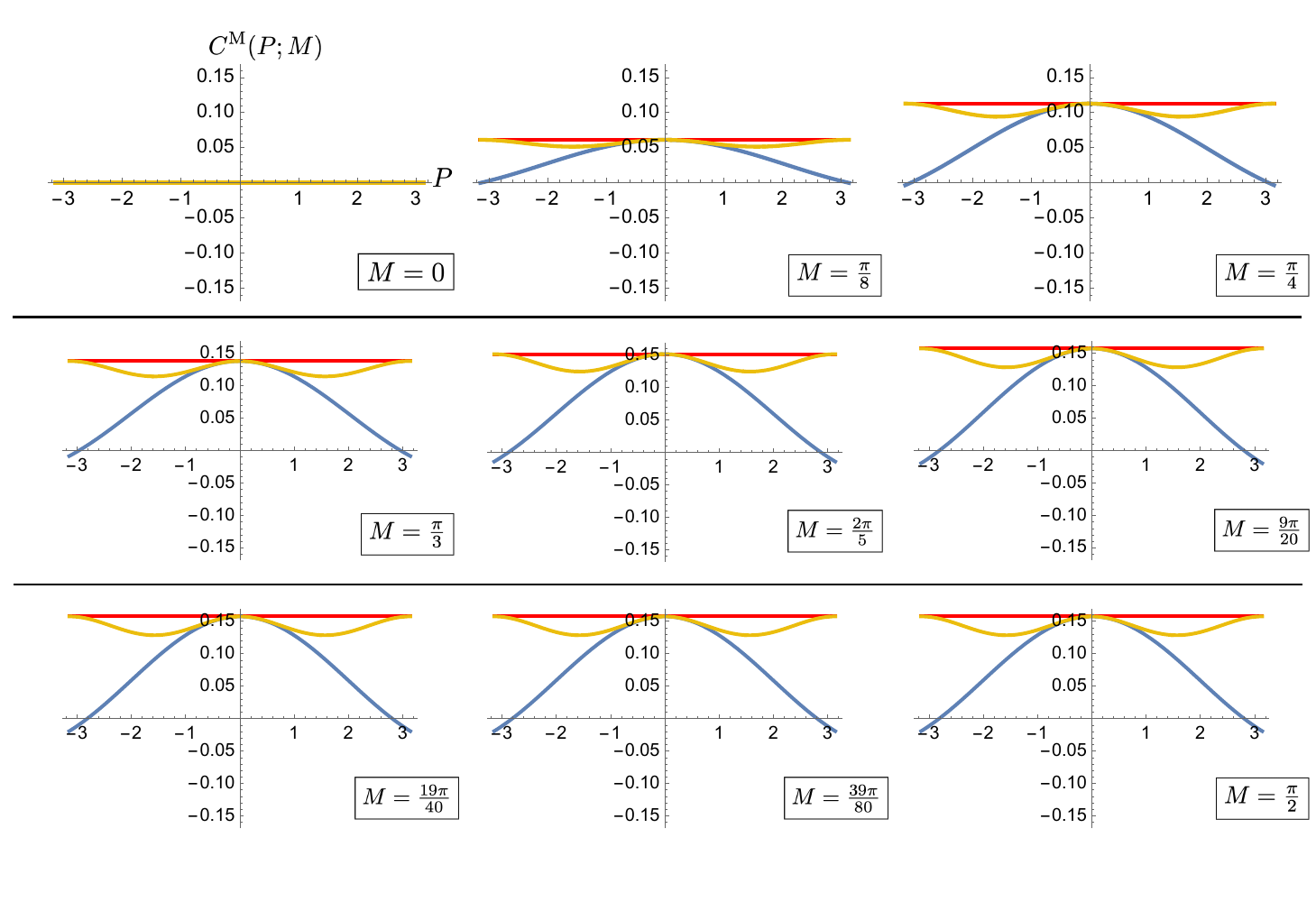}
    \caption{Same as Fig.\ \ref{fig:As} but for the $C^{\text{M}}(P;M)/(2\pi)$'s.}
    \label{fig:Cs}
\end{figure*}

\noindent
In Fig.\ \ref{fig:Bs}, we plot the $B^{\text{M}}(P;M)/(2\pi)$'s, and in \mbox{Fig.\ \ref{fig:Cs}}, we plot the $C^{\text{M}}(P;M)/(2\pi)$'s. One thing to notice for sure, is that the denominations used for the three regimes in Fig.\ \ref{fig:As} do not always match the present behaviors, which is why we have erased these denominations. Let us comment on the rest of the integration domain for the continuum model, i.e., on the fact that we have not shown the functions of $P$ for this rest of integration domain. What we said for $A^{\text{Dirac}}(P;M)/(2\pi)$ stays true for $B^{\text{Dirac}}(P;M)/(2\pi)$: the contribution of the region $]-\infty,-\pi] \cup [\pi, +\infty[$ in the momentum integral is negligible because of relatively fast asymptotics towards \emph{approximate periodicity around $0$} of the integrand at $\pm \infty$. Indeed, for $B^{\text{Dirac}}(P;M)/(2\pi)$, we have a sinc function (cardinal sinus)---rather than a $\cos$ or, rather, \mbox{a $\sin$---, so} at $\pm \infty$ the integrand of the continuum model does not go asymptotically towards \emph{exact} periodicity around $0$ at $\pm \infty$, but there is still an \emph{approximate} asymptotic periodicity around exactly $0$ that is reached relatively fast as we increase $|P|$, and, in addition, with an absolute value of the amplitude that is extremely small compared to the maximum of the function at $P=0$. Hence, overall it is not difficult to convince oneself that again the contribution of the region $]-\infty,-\pi] \cup [\pi, +\infty[$ in the momentum integral is negligible

\section{FDQCA GF as a function of the internal matrix elements of the QCA GF: proof by power expansion}
\label{app:theproof}

\noindent
In this appendix, we prove the validity of \mbox{Eq.\ \eqref{eq:thefinalalternateexpression}} by showing that $G_{\text{n.d.}}^{\text{FQW}}(t' + \Delta n \, \epsilon, x'+D\e;t',x')$ exactly vanishes when $\Delta n+D$ is even, and \mbox{$G_{\text{d.}}^{\text{FQW}}(t' + \Delta n \, \epsilon, x'+D\e;t',x')$} exactly vanishes when $\Delta n+D$ is odd. One interesting way of proving this is via an as explicit as sensible visualization of all the terms of $(\hat U_\epsilon^{\text{FQW}})^{\Delta n }$, and at the end of the proof we hope the reader will be able to connect at once all these terms, and navigation through them, to the mere structure of $\hat U_\epsilon^{\text{FQW}}$.

We know that, in a quantum-mechanical context, the GF is essentially equal to the propagator, and with our conventions it is strictly equal. For the FDQCA, the propagator is
\par\nobreak
{\small
\begin{align}
    K^{\text{FQW}}(t' + \Delta n \, \epsilon, x'+D\e;t',x')   \defeq \bra{x'+ D\e} (\hat U_\epsilon^{\text{FQW}})^{\Delta n} \ket{x'} \, .
\end{align}}
We split the one-time-step evolution operator into its diagonal and non-diagonal chirality parts:
\begin{align}
    &\hat U_\epsilon^{\text{FQW}} = \underbrace{\begin{bmatrix} c_\epsilon \hat{S}_{\epsilon} \otimes \sigma^1_{\text{flav.}} & 0 \\ 0 & c_\epsilon \hat{S}^{\dagger}_{\epsilon} \otimes \sigma^1_{\text{flav.}} \end{bmatrix}}_{\hat{U}_{\epsilon;\text{d.}}^{\text{FQW}}} \nonumber \\ & \ \ \ \ \ \ \ \  \ \ \ + \underbrace{\begin{bmatrix} 0 & - i s_\epsilon \otimes \mathbb I_{\text{flav.}} \\ - i s_\epsilon \otimes \mathbb I_{\text{flav.}} & 0 \end{bmatrix}}_{\hat{U}_{\epsilon;\text{n.d.}}^{\text{FQW}}} \, .  
\end{align}
Notice that $\hat{U}_{\epsilon;\text{d.}}^{\text{FQW}}$ shifts the spatial position by one lattice site (that is, $D = \pm 1$) and flips the flavor, whereas $\hat{U}_{\epsilon;\text{n.d.}}^{\text{FQW}}$ flips the chirality but induces no spatial translation ($D = 0$), and moreover preserves the flavor.

Expanding the power of the unitary as a binomial series yields:
\begin{equation}
    (\hat U_\epsilon^{\text{FQW}})^{\Delta n} = (\hat{U}_{\epsilon;\text{d.}}^{\text{FQW}} + \hat{U}_{\epsilon;\text{n.d.}}^{\text{FQW}})^{\Delta n} = \sum_{l=0}^{\Delta n} \mathcal{W}(\Delta n, l) \, ,
\end{equation}
where $\mathcal{W}(\Delta n, l)$ represents the sum of all possible operator sequences (i.e., words) consisting of exactly $l$ chirality flips (action of $\hat{U}_{\epsilon;\text{n.d.}}^{\text{FQW}}$) and $(\Delta n - l)$ spatial shifts (action of $\hat{U}_{\epsilon;\text{d.}}^{\text{FQW}}$).

Now, for the specific matrix element $\bra{x} \mathcal{W}(\Delta n, l) \ket{x'}$ to be non-zero, the sequence $\mathcal{W}(\Delta n, l)$ must produce at least one spatial displacement of $D$. Since all translations come strictly from the $(\Delta n - l)$ applications of $\hat{U}_{\text{d.}}^{\text{FQW}}$, the displacement $D$ that there might be must share the exact same parity as the number of spatial shifts applied:
\begin{equation}
    \Delta n - l \equiv D \pmod 2 \implies l \equiv \Delta n + D \pmod 2 \, .
\end{equation}

Tracking the parity of $l$ immediately yields the selection rules for the Green's function blocks:
\begin{itemize}
    \item \textbf{When $\Delta n + D$ is odd:} The number of chirality flips $l$ must be odd, producing a strictly anti-diagonal matrix in chirality space. The diagonal part, $G_{\text{d.}}^{\text{FQW}}(t' + \Delta n \, \epsilon, x'+D\e;t',x')$, is therefore structurally zero.
    \item \textbf{When $\Delta n + D$ is even:} The number of chirality flips $l$ must be even, producing a strictly diagonal matrix in chirality space. The anti-diagonal part, $G_{\text{n.d.}}^{\text{FQW}}(t' + \Delta n \, \epsilon, x'+D\e;t',x')$, is therefore structurally zero.
\end{itemize}

Finally, we must account for the flavor tensor structure. In any valid word $\mathcal{W}(\Delta n, l)$, the operator $\hat U_\epsilon^{\text{FQW}}$ applies exactly $(\Delta n - l)$ flavor flips. However, the definition of the FDQCA Green's function involves the modified time-shift operator $ \hat T_\epsilon^f \defeq \hat T_\epsilon \otimes \sigma_x$, which introduces one extra flavor flip per time step, adding $\Delta n$ extra flips overall. The total parity of flavor flips is thus:
\begin{equation}
    (\Delta n - l) + \Delta n = 2\Delta n - l \equiv l \pmod 2 \, .
\end{equation}
Because the flavor parity equals $l$, the flavor operator is exactly $\mathbb I_{\text{flav.}}$ when $l$ is even (the diagonal block $G_{\text{d.}}^{\text{FQW}}(t' + \Delta n \, \epsilon, x'+D\e;t',x')$), and $\sigma^1_{\text{flav.}}$ when $l$ is odd (the anti-diagonal block $G_{\text{n.d.}}^{\text{FQW}}(t' + \Delta n \, \epsilon, x'+D\e;t',x')$). With this, we recover the bipartite structure of Eq.\ \eqref{eq:thefinalalternateexpression}, i.e., the proof is completed. \\

\section{FDQCA GF as a function of the internal matrix elements of the QCA GF: proof by induction}
\label{app:induction}

\noindent
In this appendix, we prove Eq.\ \eqref{eq:thefinalalternateexpression} by induction. The precise statement we want to prove is: ``For any $\Delta n \in \mathbb{N}^\ast$, Eq.\ \eqref{eq:thefinalalternateexpression} holds.''.

The reader can check that the statement is true for $\Delta n = 1$, since evaluating the RHS Eq.\ \eqref{eq:thefinalalternateexpression} for $\Delta n = 1$ does coincide with the RHS of the Eq.\ \eqref{eq:thelastEQonEarth}, which is the known, fully explicit expression for $G^{\text{FQW}}(t,x;t',x')$.

We now assume that the statement is true for an arbitrary $\Delta n$. Let us investigate the expression of $G^{\text{FQW}}(\tau,x;t',x')$ for $\tau \defeq (\Delta n +1)\epsilon$:
\begin{subequations}
\begin{align}
&G((\Delta n +1)\epsilon + t',x;t',x') \nonumber\\
& \ \ \ = \bra{x}  (\hat U^{\text{FQW}}_\e)^{\Delta n +1} \ket{x'} \\
& \ \ \ = \bra{x} \hat U^{\text{FQW}}_\e (\hat U^{\text{FQW}}_\e)^{\Delta n} \ket{x'} \\
& \ \ \ = \sum_{y\in\mathbb Z \e} \bra{x} \hat U^{\text{FQW}}_\e \ket{y} \bra{y} (\hat U^{\text{FQW}}_\e)^{\Delta n} \ket{x'} \, .
\end{align}
\end{subequations}
Now, using the RHS of Eq.\ \eqref{eq:thelastEQonEarth} for $\bra{x} \hat U^{\text{FQW}}_\e \ket{y}$, and using the induction hypothesis for $\bra{y} (\hat U^{\text{FQW}}_\e)^{\Delta n} \ket{x'}$, the previous equation reads
{\par\nobreak
\small
\begin{align}
&G((\Delta n +1)\epsilon + t',x;t',x') \nonumber \\
& \  = \sum_{y\in\mathbb Z \e}
\frac{1}{\e} \bigg\{ 
\begin{bmatrix}
    c_\e \delta_{y,x-\e} & 0 \\ 0 &  c_\e \delta_{y,x+\e}
\end{bmatrix} \otimes \mathbb I_{\text{flav.}}    \nonumber \\ 
&\ \ \ \ \ \  \ \ \ \ \ \ \ \   + \Big( (-is_\e) \delta_{y,x} \, \sigma^1_{\text{chir.}} \Big) \otimes \sigma^1_{\text{flav.}} 
\bigg\} \nonumber  \\
&\ \ \ \ \ \ \ \ \ \  \times \left\{ 
\frac{1}{2} \Big[ (-1)^{D+\Delta n} +1 \Big] G^{\text{QW}}_{\text{d.}}(t,x;t',x') \otimes \mathbb I_{\text{flav.}} \right. \nonumber \\ 
& \ \ \ \ \ \ \ \ \ \ \ \ \ \  \left. + \frac{1}{2} \Big[ (-1)^{D+\Delta n +1} +1 \Big] G^{\text{QW}}_{\text{n.d.}}(t,x;t',x') \otimes \mathbb \sigma^1_{\text{flav.}}
\right\}
\end{align}}
After several crucial but straightforward lines of computation, we find the following expression, which, first of all, is compatible with the result we want to prove:
\begin{equation}
G((\Delta n +1)\epsilon + t',x;t',x') = A + B \, ,    
\end{equation}
where
\begin{align}
&A = \Big[ (-1)^{D+(\Delta n +1)} +1 \Big] \nonumber \\  
&\ \ \ \  \times 
\begin{bmatrix}
     F^{\text{QW}}_{11}(t,x;t',x')& 0  \\
0 & F^{\text{QW}}_{22}(t,x;t',x')
\end{bmatrix} \otimes \mathbb I_{\text{flav.}} \, ,
\end{align}
with
{\par\nobreak
\small
\begin{subequations}
\begin{align}
F^{\text{QW}}_{11}(t,x;t',x') &\defeq -is_\e G^{\text{QW}}_{21}(t,x;t',x') + c_\e G^{\text{QW}}_{11}(t,x-\e;t',x') \\
F^{\text{QW}}_{22}(t,x;t',x') &\defeq  -is_\e G^{\text{QW}}_{12}(t,x;t',x') + c_\e G^{\text{QW}}_{22}(t,x+\e;t',x') \, ,
\end{align}    
\end{subequations}}
and
\begin{align}
&B = \Big[ (-1)^{D+(\Delta n +1) + 1} +1 \Big] \nonumber \\  
&\ \ \ \  \times 
\begin{bmatrix}
    0 & F^{\text{QW}}_{12}(t,x;t',x')  \\
F^{\text{QW}}_{21}(t,x;t',x') & 0
\end{bmatrix} \otimes \sigma^1_{\text{flav.}} \, ,
\end{align}
with
{\par\nobreak
\small
\begin{subequations}
\begin{align}
F^{\text{QW}}_{12}(t,x;t',x') \defeq -is_\e G^{\text{QW}}_{22}(t,x;t',x') + c_\e G^{\text{QW}}_{12}(t,x-\e;t',x') \\
F^{\text{QW}}_{21}(t,x;t',x') \defeq  -is_\e G^{\text{QW}}_{11}(t,x;t',x') + c_\e G^{\text{QW}}_{21}(t,x+\e;t',x') \, .
\end{align}    
\end{subequations}}

The careful reader may have noticed that there is just one equality left to prove, namely, 
\begin{equation}
\label{eq:remainingeqtoprove}
 F^{\text{QW}}(t,x;t',x') = G^{\text{QW}}(t+\epsilon,x;t',x') \, ,
\end{equation}
which of course is a matrix equality. And this equality can be shown to be true by having in mind that that
\begin{subequations}
\begin{align}
&G^{\text{QW}}(t+\e,x;t',x') \\
& \ \ \ \ = \sum_{y\in\mathbb Z \e} \bra{x} \hat U^{\text{QW}}_\e \ket{y} \bra{y} (\hat U^{\text{QW}}_\e)^{\Delta n} \ket{x'} \\ 
& \ \ \ \  = \sum_{y\in\mathbb Z \e} G^{\text{QW}}(t+\e,x;t,y) G^{\text{QW}}(t,x;t',x') \, ,
\end{align}
\end{subequations}
where $G^{\text{QW}}(t+\e,x;t,y)$ is given by Eq.\ \eqref{eq:thefinaleqisthisone} and $G^{\text{QW}}(t,x;t',x')$ simply keeps an non-explicit form, so that, after doing the matrix multiplication and after eliminating the $y$ variable, we do obtain \mbox{Eq.\ \eqref{eq:remainingeqtoprove}}, which completes our proof by induction.

\section{Computing the one-time-step FDQCA Green's function}
\label{app:computGdiagandnondiag}

\noindent
There are usually two ways of computing a direct-space GF in a quantum-mechanical context. The first is the standard one of GF theory for partial differential equations, and corresponds to finding first the spacetime Fourier transform of the GF, and to then integrating it over the reciprocal-space domain. The second is to remember that, in quantum mechanics, the GF is essentially equal to the so-called propagator, and to compute the propagator via usual QM methods; for a free theory, the usual method consists in inserting a momentum closure relation in the basic expression of the propagator, because this momentum-diagonalizes the Hamiltonian and the evolution operator since in a free theory these operators only depend on the momentum operator.

In the case of the original QCA, we have worked out the two methods, and checked that they coincided. Here, for the FDQCA, the situation is more complicated regarding the first method. Indeed, the fact that $\Delta n =1$ changes nothing to the problem we encountered in \mbox{Sec.\ \ref{subsubsec:findingthesimplestexpression}}, which is that we were only able to compute parts of the Green's function by Fourier integration, but not the whole. Hence, one could turn to the second method, and indeed it works, one simply has to do a computation analog to that performed for the QCA/QW in App.\ \ref{app:propagator}.

Now, in our QCA/QW context, there is another, third way of computing the GF, which we want to mention, since it is actually the simplest one in that context. This method is similar to the second method mentioned previously, but without inserting any momentum closure relation. This method is the simplest one here because, although the one-time-step evolution operator of the FQW depends (solely) on the momentum operator as for any free theory, since this dependence is via translation operators \emph{always}, the action of the whole one-time-step evolution operator on position kets is trivial, we do not need to go to spatial Fourier space.

This in the end delivers
\begin{equation}
\label{eq:thelastEQonEarth}
G^{\text{FQW}}(t'+\epsilon,x;t',x') = \frac{1}{\e} 
\begin{bmatrix}
c_{\e} \delta_{x',x-\e} \, \mathbb   I_{\text{flav.}} & -is_\e \delta_{x',x} \,  \sigma^1_{\text{flav.}} \\
-is_\e \delta_{x',x} \,\sigma^1_{\text{flav.}}  &   c_{\e} \delta_{x',x+\e}    \, \mathbb   I_{\text{flav.}}
\end{bmatrix} \, .
\end{equation}

\end{document}